\documentclass[10pt, a4paper, twocolumn, aps, pra, showpacs, superscriptaddress, nofootinbib]{revtex4-1}
\pdfoutput=1
\usepackage[utf8x]{inputenc}
\usepackage{microtype}
\usepackage{amsmath}
\usepackage{amsfonts}
\usepackage{amssymb}
\usepackage{cellspace,booktabs,array}
\usepackage{hyperref}
\usepackage[dvipsnames]{xcolor}
\usepackage{graphicx}
\usepackage{siunitx}
\usepackage{tabularx}

\usepackage{physics}

\usepackage{bbold}

\begin{document}

\title{Quantum interference device for controlled two-qubit operations}

\begin{abstract}
    Universal quantum computing relies on high-fidelity entangling operations. Here we demonstrate that four coupled qubits can operate as a quantum gate, where two qubits control the operation on two target qubits (a four-qubit gate). This configuration can implement four different controlled two-qubit gates: two different entangling swap and phase operations, a phase operation distinguishing states of different parity, and the identity operation (idle quantum gate), where the choice of gate is set by the state of the control qubits.
    The device exploits quantum interference to control the operation on the target qubits by coupling them to each other via the control qubits.
    By connecting several four-qubit devices in a two-dimensional lattice, one can achieve a highly connected quantum computer. We consider an implementation of the four-qubit gate with superconducting qubits, using capacitively coupled qubits arranged in a diamond-shaped architecture.
\end{abstract}

\date{\today}
\author{Niels Jakob Søe Loft}
\thanks{Correspondence: Niels Jakob Søe Loft (nsl@phys.au.dk)}
\affiliation{Department of Physics and Astronomy, Aarhus University, DK-8000 Aarhus C, Denmark}
\author{Morten Kjaergaard}
\affiliation{Research Laboratory of Electronics, Massachusetts Institute of Technology, Cambridge, MA 02139, USA}
\author{Lasse Bjørn Kristensen}
\affiliation{Department of Physics and Astronomy, Aarhus University, DK-8000 Aarhus C, Denmark}
\author{Christian Kraglund Andersen}
\affiliation{Department of Physics, ETH Zurich, CH-8093 Zurich, Switzerland}
\author{Thorvald W. Larsen}
\affiliation{Center for Quantum Devices and Microsoft Quantum Lab--Copenhagen, Niels Bohr Institute, University of Copenhagen, DK-2100 Copenhagen, Denmark}
\author{Simon Gustavsson}
\affiliation{Research Laboratory of Electronics, Massachusetts Institute of Technology, Cambridge, MA 02139, USA}
\author{William D. Oliver}
\affiliation{Research Laboratory of Electronics, Massachusetts Institute of Technology, Cambridge, MA 02139, USA}
\affiliation{MIT Lincoln Laboratory, 244 Wood Street, Lexington, MA 02420, USA}
\affiliation{Department of Physics, Massachusetts Institute of Technology, Cambridge, MA 02139, USA}
\affiliation{Department of Electrical Engineering and Computer Science, Massachusetts Institute of Technology, Cambridge, MA 02139, USA}
\author{Nikolaj T. Zinner}
\affiliation{Department of Physics and Astronomy, Aarhus University, DK-8000 Aarhus C, Denmark}
\affiliation{Aarhus Institute of Advanced Study, Aarhus University, DK-8000 Aarhus C, Denmark}

\maketitle

\section{Introduction}

The goal of quantum computing is to implement a programmable quantum information processor. Such a processor requires access to a universal gate set from which any quantum algorithm can be constructed. Universal gate sets can be formed from single-qubit gates supplemented by a two-qubit entangling gate\cite{divincenzo1995universalqc}.
Furthermore, fault-tolerance is necessary in order to perform arbitrarily long and precise computations, which, for the most lenient error correcting surface codes, puts a lower bound of around 0.99 on the required gate fidelities\cite{raussendorf2007, corcoles2015, obrien2017, fowler2012}.
Extensible high-fidelity entangling two-qubit gates are thus key elements in any multi-purpose quantum information processor.

Single-qubit gate operations are routinely performed with fidelities above 0.99\cite{buluta2011review, gustavsson2013, reagor2018rigetti, rol2017, sheldon2016, chen2016, barends2014cz, wright2019, harty2014, itoh2014, zajac2018}, but pushing two-qubit gate fidelities above 0.99 still proves a daunting task.
Despite the challenges in realizing a low loss environment while at the same time having high control of two-qubit operations, several two-qubit gates have been reported to do so.
The first group to accomplish this was Benhelm \textit{et al.}, who in 2008 demonstrated a Mølmer-Sørensen-type entangling gate\cite{sorensen1999, sorensen2000} with a fidelity of 0.993 using laser-controlled trapped calcium ions\cite{benhelm2008}.
Since then, similar ion trap experiments have realized high-fidelity two-qubit gates\cite{gaebler2016, ballance2016, erhard2019, harty2016, ballance2015}.
Another promising qubit architecture is silicon-based quantum dots\cite{koiller2001, friesen2003, itoh2014, zajac2018}, where controlled-rotation gates were recently benchmarked with a fidelity of 0.98\cite{huang2019}.

%
%
%
%
%
%
%

In superconducting qubits the controlled-phase (\textsf{CZ}) gate~\cite{barends2014cz, kelly2014cz, Yu2014, rol2019, kjaergaard2019} and the cross-resonance (\textsf{CR}) gate\cite{Sheldon2016crgate} have been shown to exceed a fidelity of 0.99.
Other two-qubit gates, like the $i\textsf{SWAP}$ and $\sqrt{i\textsf{SWAP}}$ gates\cite{mckay2016iswap, Dewes2012sqrtiswap, salathe2015iswap, reagor2018rigetti, caldwell2017}, $b\textsf{SWAP}$ gate\cite{poletto2012bswap}, the resonator induced phase (\textsf{RIP}) gate\cite{paik2016ripgate}, and a parametric \textsf{CZ} gate\cite{reagor2018rigetti, caldwell2017}, have been demonstrated with fidelities in the 0.9's.
These quantum gates are typically performed with transmons\cite{transmon_original, you2007pretransmon, Barends2013xmon, Yu2014}, coupled directly to each other or via a separate coupling element, e.g. a transmission line resonator or a tunable coupler.

In this work, we propose the implementation of controlled two-qubit operations utilizing quantum interference patterns in a network of four qubits.
As a specific architecture, where this four-qubit gate can be implemented natively, we consider superconducting transmon qubits placed in a diamond-shaped geometry.
The qubits are coupled only through simple capacitive couplings. A similar 2D array of transmons was considered in Refs.~\cite{lloyd2016, ciani2019}, but with different couplings and purpose.
The system comprises a four-qubit quantum gate (`the diamond gate'), where the state of two qubits control a two-qubit gate operation on the remaining two qubits.
Since the diamond gate natively implements multiple unitaries, it is a useful addition to the gate set used for quantum simulation and quantum compilation.
Due to its ability to perform (controlled) two-qubit entangling operations, supplementing the diamond gate with single-qubit operations allows for universal quantum computing on the target qubits.

In Sec.~\ref{sec:system} we discuss the operation of the diamond gate, and in Sec.~\ref{sec:extensibility} how it can constitute a building block in an extensible quantum computer.
In Sec.~\ref{sec:simulations} we simulate the transmon implementation of the gate, using parameters from state-of-the-art superconducting qubits, in a Lindblad master equation simulation.
We find that the gate generally operates with fidelity around 0.99 in less than 100 ns.
Finally, in Sec.~\ref{sec:qutrits} we consider the effects of couplings to higher-energy states in the transmon spectrum, leading to undesired leakage across the control. We show how this behavior can be counteracted by engineering a cross-coupling to cancel the effects.
This is a passive scheme, in contrast to the microwave pulse-based scheme recently shown to reduce leakage in the \textsf{CZ} gate\cite{rol2019}.

Throughout this paper, we use units where $\hbar = 1$.

\section{Results}

\subsection{Four-qubit diamond gate}
\label{sec:system}

\begin{figure*}
  \includegraphics[width=\textwidth]{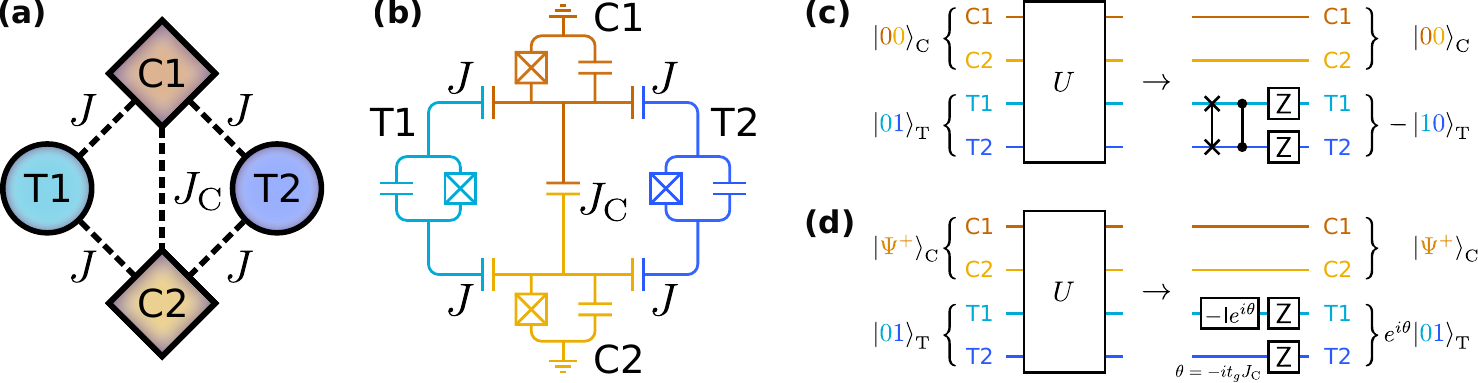}
  \caption{
  \textbf{(a)} The diamond gate: Four-qubit system consisting of two target qubits (T1 and T2) and control qubits (C1 and C2) coupled through exchange interactions (dashed lines) with the indicated strengths.
  \textbf{(b)} Lumped element superconducting circuit diagram of four capacitively coupled transmons, where each colored subcircuit corresponds to the same-colored qubit in \textbf{(a)}.
  \textbf{(c)--(d)} Example transformations implemented by the diamond gate, $U$, of Eq.~\eqref{eq:U}.}
  \label{fig:superconducting_circuit}
\end{figure*}

Consider the four-qubit Hamiltonian being a sum of the non-interacting part
\begin{equation}
   H_0 = -\frac{1}{2} (\Omega + \Delta) (\sigma_z^\text{T1} + \sigma_z^\text{T2})
   - \frac{1}{2} \Omega (\sigma_z^\text{C1} + \sigma_z^\text{C2}) \; ,
   \label{eq:H0}
\end{equation}
where $\Omega + \Delta$ ($\Omega$) is the fixed frequency of the target (control) qubits, and the interaction terms
\begin{equation}
   H_\text{int} = J_\text{C} \, \sigma_y^\text{C1}\sigma_y^\text{C2}
   + J (\sigma_y^{\text{T}1} + \sigma_y^{\text{T}2}) (\sigma_y^{\text{C}1} + \sigma_y^{\text{C}2}) \; .
   \label{eq:Hint}
\end{equation}
Here $\sigma_z^j = \dyad{0}{0}_j - \dyad{1}{1}_j$ and $\sigma_y^j = i\dyad{1}{0}_j - i\dyad{1}{0}_j$ are Pauli operators on qubit $j$, and the qubit frequencies are assumed positive such that $\ket{0}_j$ is the non-interacting qubit ground state. For simplicity we have assumed that the two target (control) qubits are on resonance, and that all the couplings between the target and control qubits have the same strength $J$, although, as we will show later, this contraint is not needed for high performance of the gate. The four-qubit system is sketched in Figure~\ref{fig:superconducting_circuit}a.
As we will discuss in the following, the system implements a four-qubit gate, which we will refer to as `the diamond gate' due to the geometry of the system.

Superconducting circuits offer a natural platform for implementing this type of Hamiltonian\cite{krantz2019}. Specifically, by truncating the Hilbert space for each degree of freedom to qubits, the circuit of four capacitively coupled transmon qubits in Figure~\ref{fig:superconducting_circuit}b implements the Hamiltonian.
Later, we analyze the model including the second excited state of the transmon qubits.

We now consider the interaction Hamiltonian, $H_\text{int}$, in the frame rotating with $H_0$ and simplify the expression by assuming $\lvert 2\Omega \rvert \gg \lvert J \rvert$ (rotating wave approximation), which allows us to ignore the most rapidly oscillating terms. The system Hamiltonian is then
\begin{equation}
   H = J_\text{C} \, \sigma_+^\text{C1}\sigma_-^\text{C2}
   + J \, e^{i\Delta t}
   (\sigma_+^{\text{T}1} + \sigma_+^{\text{T}2}) (\sigma_-^{\text{C}1} + \sigma_-^{\text{C}2})
   + \text{H.c.} \; ,
   \label{eq:H}
\end{equation}
with $\sigma_+^j = \dyad{1}{0}_j$ and $\sigma_-^j = \dyad{0}{1}_j$ on qubit $j$. This Hamiltonian governs the dynamics resulting from the interactions in the model. We show in Appendix~\ref{sec:unitary_gate_analysis} that the effective unitary time-evolution of $H$ gives rise to a four-qubit gate operating by means of controlled quantum interference (the diamond gate).
The analysis is based on a Magnus expansion of $H$ within Floquet theory, which assumes $\lvert \Delta \rvert \gg \lvert J \rvert, \lvert J_\text{C} \rvert$, i.e. a qubit detuning much larger than the coupling strengths.

The diamond gate is a four-way controlled two-qubit gate operation on the target qubits T1 and T2.
Consider the following gates in the target qubit computational basis, $\{ \ket{00}_\text{T}, \ket{01}_\text{T}, \ket{10}_\text{T}, \ket{11}_\text{T} \}$, where the superscripts refer to the control setting (discussed below):
\begin{align}
    \label{eq:UT00}
    U^{00}_\text{T} ={} &
    \begin{pmatrix}
        \makebox[1.1em]{$1$} & 0 & 0 & 0 \\[.2em]
        0 & 0 & \makebox[1.1em]{$-1$} & 0 \\[.2em]
        0 & \makebox[1.1em]{$-1$} & 0 & 0 \\[.2em]
        0 & 0 & 0 & \makebox[1.1em]{$-1$}
    \end{pmatrix}
    = \textsf{ZZ} \cdot \textsf{CZ} \cdot \textsf{SWAP}
    \; , \\
    \label{eq:UT11}
    U^{11}_\text{T} ={} &
    \begin{pmatrix}
        \makebox[1.1em]{$-1$} & 0 & 0 & 0 \\[.2em]
        0 & 0 & \makebox[1.1em]{$-1$} & 0 \\[.2em]
        0 & \makebox[1.1em]{$-1$} & 0 & 0 \\[.2em]
        0 & 0 & 0 & \makebox[1.1em]{$1$}
    \end{pmatrix}
    = - \textsf{CZ} \cdot \textsf{SWAP}
    \; , \\
    \label{eq:UTpl}
    U^{\Psi^+}_\text{T} ={} &
    \begin{pmatrix}
        \makebox[1.1em]{$-1$} & 0 & 0 & 0 \\[.2em]
        0 & \makebox[1.1em]{$1$} & 0 & 0 \\[.2em]
        0 & 0 & \makebox[1.1em]{$1$} & 0 \\[.2em]
        0 & 0 & 0 & \makebox[1.1em]{$-1$}
    \end{pmatrix}
    e^{-it_g J_\text{C}}
    = - \textsf{ZZ} \, e^{-it_g J_\text{C}}
    \; , \\
    \label{eq:UTmi}
    U^{\Psi^-}_\text{T} ={} &
    \begin{pmatrix}
        \makebox[1.1em]{$1$} & 0 & 0 & 0 \\[.2em]
        0 & \makebox[1.1em]{$1$} & 0 & 0 \\[.2em]
        0 & 0 & \makebox[1.1em]{$1$} & 0 \\[.2em]
        0 & 0 & 0 & \makebox[1.1em]{$1$}
    \end{pmatrix}
    e^{+it_g J_\text{C}}
    = \textsf{II} \, e^{+it_g J_\text{C}}
    \; .
\end{align}
Here $t_g$ is the gate time given by
\begin{equation}
   t_g = \frac{\pi\lvert \Delta\rvert}{4J^2} \; .
   \label{eq:tg}
\end{equation}
Eqs.~\eqref{eq:UT00}--\eqref{eq:UTmi} show the two-qubit operations in terms of well-known gates from the literature, see e.g. Ref.~\cite{nielsen_chuang_2010}. Here $\textsf{ZZ}$ is understood as a $\textsf{Z}$ gate on each target qubit. Thus we see that $U^{00}_\text{T}$ and $U^{11}_\text{T}$ are two different combined swap and phase operations.
Access to just one of these entangling gates will facilitate universal quantum computing.
The third gate, $U^{\Psi^-}_\text{T}$, is a phase operation distinguishing target states with different parity (addition of T1 and T2's bit value modulo 2) by application of a relative sign.
The final gate, $U_\text{T}^{\Psi^-}$, which just adds a global phase, is the identity gate. We can therefore regard the preceding three gates as actual computational gates, while $U_\text{T}^{\Psi^-}$ is the idle position of the device.

The above two-qubit gates are controlled by the state of the control qubits, which we describe in the following orthonormal basis: $\{ \ket{00}_\text{C}, \ket{11}_\text{C}, \ket{\Psi^+}_\text{C}, \ket{\Psi^-}_\text{C} \}$.
We refer to this basis, which mixes computational basis states and the Bell states $\ket{\Psi^\pm}_\text{C} = (\ket{01}_\text{C} \pm \ket{10}_\text{C})/\sqrt 2$, as the control basis. The full four-qubit unitary operation of the diamond gate is
\begin{equation}
    \label{eq:U}
    \begin{aligned}
        U ={} &\dyad{00}_\text{C} U^{00}_\text{T} + \dyad{11}_\text{C} U^{11}_\text{T} \\
              &+ \dyad{\Psi^+}_\text{C} U^{\Psi^+}_\text{T} + \dyad{\Psi^-}_\text{C} U^{\Psi^-}_\text{T} \; .
    \end{aligned}
\end{equation}
Cast this way, it is evident that $U$ describes a four-way controlled operation on the target qubits. If the control qubits are initialized in one of the control basis states, only the corresponding gate among \eqref{eq:UT00}--\eqref{eq:UTmi} is performed. The control state is unchanged after the gate operation.
Figure~\ref{fig:superconducting_circuit}c--d illustrate the gate operation on the target state $\ket{01}_\text{T}$ in the cases where the control is $\ket{00}_\text{C}$ and $\ket{\Psi^+}_\text{C}$, respectively. However, these gate diagrams only show the gate operation for these two control states, and in general the diamond gate performs a unitary operation on any initial four-qubit state.
A more sophisticated decomposition of the full unitary $U$ is given i Figure~\ref{fig:gate_circuit} in Appendix~\ref{sec:unitary_gate_analysis}, where we note that the complexity in terms of number of $\textsf{CNOT}$ gates is 42.
Have access to four controlled two-qubit operations natively is useful for quantum simulation and may ease quantum gate compilation significantly.

As shown in Appendix~\ref{sec:unitary_gate_analysis}, the unitary time-evolution under the Hamiltonian of Eq.~\eqref{eq:H} approximately gives rise to $U$. Within the first order Magnus expansion, the approximation is exact when $J_\text{C} = 0$, however a non-zero coupling between the control qubits is needed in order to initialze the control Bell states.
Such a coupling allows the triplet states $\{ \ket{00}_\text{C} , \ket{11}_\text{C}, \ket{\Psi^+}_\text{C} \}$ to mix slightly during the gate operation, in which case the separation of control states in Eq.~\eqref{eq:U} is no longer exact.
This leads to small gate infidelities of the order $(2J/\Delta)^2 = \pi/(t_g\Delta)$ when then control qubits are initialized in $\ket{00}_\text{C}$ or $\ket{11}_\text{C}$, and twice as large when the control is in $\ket{\Psi^+}_\text{C}$.
For typical superconducting circuit parameter values, like the ones used in the following section, these infidelities are on the order $10^{-3}$ to $10^{-2}$.
Notice that the infidelity scales inversely with the gate time, leading to a trade-off between a fast gate and high-fidelity coherent operations.
Since the singlet state $\ket{\Psi^-}_\text{C}$ does not mix with the triplet states, the idle gate operation is not affected by the coupling $J_\text{C}$, and the gate fidelity is only limited by other factors, e.g. qubit decoherence.

As mentioned above, the performance of the gate is increased if $J_\text{C} = 0$, however a non-zero direct coupling between the control qubits is necessary if we wish to preparate the entangled Bell states.
In the following, we will assume a fixed value of $J_\text{C}$, although ideally a tunable coupler\cite{yan2018} can be used to turn on the coupling only during control state preparation.
If the control qubits are detuned from the target qubits, $\lvert \Delta \rvert \gg \lvert J \rvert$, we can initialize the control state without affecting the target qubits.
This detuning can be achieved by flux tunable devices, or by fabricating single-junction qubits with different frequencies.
Thus, ignoring the oscillating terms of Eq.~\ref{eq:H}, we have effectively decoupled the control and target qubits.
We note that the effective Hamiltonian of the control qubits in the rotating frame, $J_\text{C} (\sigma_+^\text{C1}\sigma_-^\text{C2} + \sigma_-^\text{C1}\sigma_+^\text{C2})$, has a zero-energy subspace spanned by $\ket{00}_\text{C}$ and $\ket{11}_\text{C}$, and eigenstates $\ket{\Psi^\pm}_\text{C}$ of energy $\pm J_\text{C}$.
An energy separation of $J_\text{C}/2\pi \sim \SI{20}{\mega\hertz}$ allows us to initialize the control in $\ket{\Psi^\pm}_\text{C}$ by driving energy transitions\cite{poletto2012bswap, Sheldon2016crgate}.
To initialize the control in $\ket{00}_\text{C}$ or $\ket{11}_\text{C}$, we can induce Rabi oscillations between these two states by driving the control qubits similarly to the procedure analyzed in Ref.~\cite{didier2018}.

\subsection{Extensible quantum computer}
\label{sec:extensibility}

\begin{figure}
    \includegraphics[width=\columnwidth]{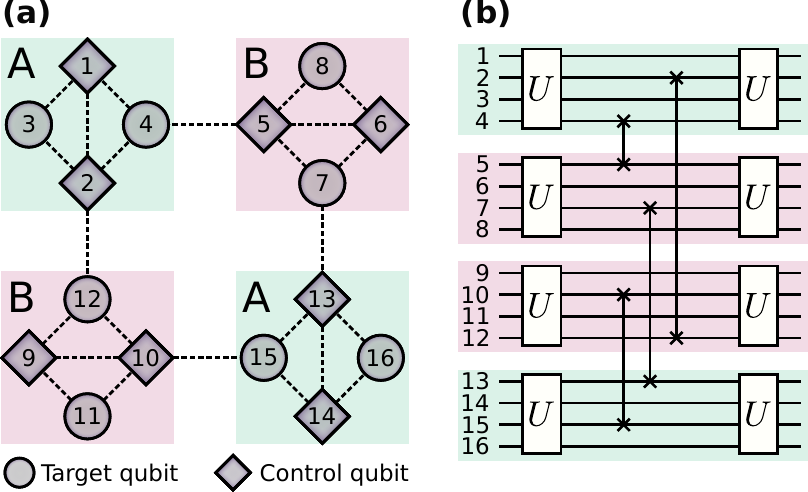}
    \caption{Proposed architecture for an extensible quantum computer. \textbf{(a)} Four connected copies of the four-qubit diamond gate device. Detuning the qubits on the plaquettes A from the qubits on the plaquettes B allows each four-qubit device to run the diamond gate independently, while tuning the connecting qubits into resonance allows swap operations between plaquettes A and B. \textbf{(b)} A sequence of diamond gates $U$ of Eq.~\eqref{eq:U} in each plaquette and two-qubit swaps between the plaquettes running on the 16-qubit quantum computer.}
    \label{fig:extensibility}
\end{figure}

The four-qubit quantum interference device can constitute a building block in an extensible quantum computer by connecting several copies.
One possible architecture is illustrated in Figure~\ref{fig:extensibility}a, where a 16-qubit quantum computer is constructed by connecting four copies of the four-qubit device, for instance through capacitive couplings.
On the plaquettes labelled A the control qubits are oriented vertically (1, 2, 13 and 14) and the target qubits horizontally (3, 4, 15 and 16), while the diamond gate devices on the plaquettes B are rotated by ninety degrees, such that control and target qubits from different plaquettes are connected.
This design of alternating A and B plaquettes can be extended in a straight-forward manner in one or two dimensions.

The quantum algorithm shown in Figure~\ref{fig:extensibility}b is a generic algorithm spreading entanglement in the computer. Supplemented with single-qubit rotations, it may serve as a variational quantum eigensolver\cite{Abhinav2017ibm}.
The algorithim can be implemented in the following way. Initially, the plaquette A qubits are far detuned from the plaquette B qubits, allowing each four-qubit diamond gate device to run the unitary gate $U$ of Eq.~\eqref{eq:U} independently. After the completion of the gates, we can prevent further dynamics within each plaquette by switching the controls to the idle state. Then, by tuning pairs of connected qubits from different plaquettes into resonance, for instance 4 and 5, we can perform swap gates or use a suitable microwave driving to perform other desired two-qubit operations.
Finally, by tuning the qubits out of resonance, and potentially switching certain controls, we are ready to run the diamond gate again.

\subsection{Numerical simulations}
\label{sec:simulations}

Although the analytic results suggest a functioning four-qubit diamond gate, we use numerical simulations to quantify the performance of the gates for state-of-the-art superconducting qubit parameters\cite{wang2018, kounalakis2018, Wendin2016cqed_review}. Decoherence is included via the Lindblad master equation,
\begin{equation}
  \dot \rho = - i [H, \rho]
  + \sum_n \Big[ C_n \rho C_n^\dagger
  - \frac{1}{2} (\rho C_n^\dagger C_n +  C_n^\dagger C_n \rho)
  \Big] \; .
  \label{eq:master-equation}
\end{equation}
Here $\rho$ is the density matrix, $H$ is the Hamiltonian of Eq.~\eqref{eq:H}, and the sum is taken over the following eight collapse operators, $C_n$: $\sqrt{\gamma} \, \sigma_z^{i}$ inducing pure dephasing and $\sqrt{\gamma} \, \sigma_-^{i}$ inducing qubit relaxation (photon loss), with $i$ running over all four qubits, denoting by $\gamma$ the decoherence rate. We solve the master equation numerically using the Python toolbox QuTiP\cite{qutip}.

\begin{table}
    \begin{tabular}{lll}
        \toprule
        & Parameter set 1 & Parameter set 2 \\
        \midrule
        $J_\text{C} / 2\pi \, \si{\mega\hertz}$ & $20$ & $20$ \\
        $J / 2\pi \, \si{\mega\hertz}$          & $65$ & $45$ \\
        $\Delta / 2\pi \, \si{\giga\hertz}$     & $2$  & $0.5$ \\
        $\gamma / \si{\mega\hertz}$             & $0.01$ & $0.01$ \\
        \midrule
        Predicted $t_g / \si{\nano\second}$     & $59.2$ & $30.9$ \\
        Simulated $t_g / \si{\nano\second}$     & $59.3$ & $31.5$ \\
        \midrule
        $F_{00}(t_g)$          & $0.9943$ & $0.9662$ \\
        $F_{11}(t_g)$          & $0.9931$ & $0.9668$ \\
        $F_{\Psi^+}(t_g)$      & $0.9881$ & $0.9348$ \\
        $F_{\Psi^-}(t_g)$      & $0.9968$ & $0.9983$ \\
        $F(t_g)$               & $0.9923$ & $0.9637$ \\
        \bottomrule
    \end{tabular}
    \caption{Two sets of model parameters and their corresponding gate times and gate fidelities. The gate fidelities are found at the simulated $t_g$.}
    \label{tab:parameters}
\end{table}

\begin{figure}
    \includegraphics[width=\columnwidth]{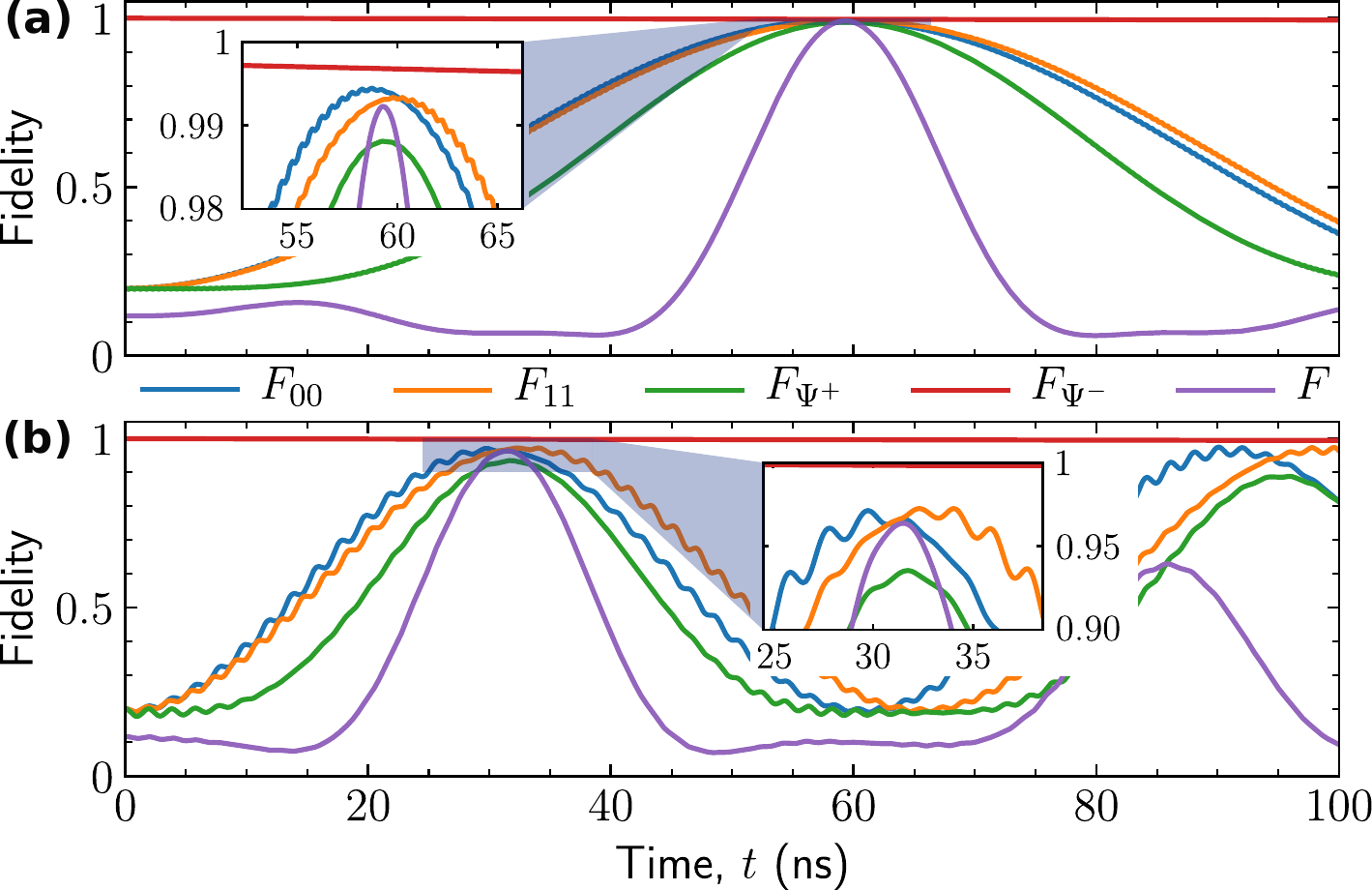}
    \caption{Fidelities versus time for the individually controlled gates ($F_{00}$, $F_{11}$, $F_{\Psi^+}$, $F_{\Psi^-}$) and the total diamond gate ($F$). Insets show zooms around the gate time.
    The parameters used in \textbf{(a)} are set 1 from Table~\ref{tab:parameters}, and in \textbf{(b)} they are set 2.}
    \label{fig:fid_vs_time}
\end{figure}

\begin{figure*}[t]
   \includegraphics[width=2\columnwidth]{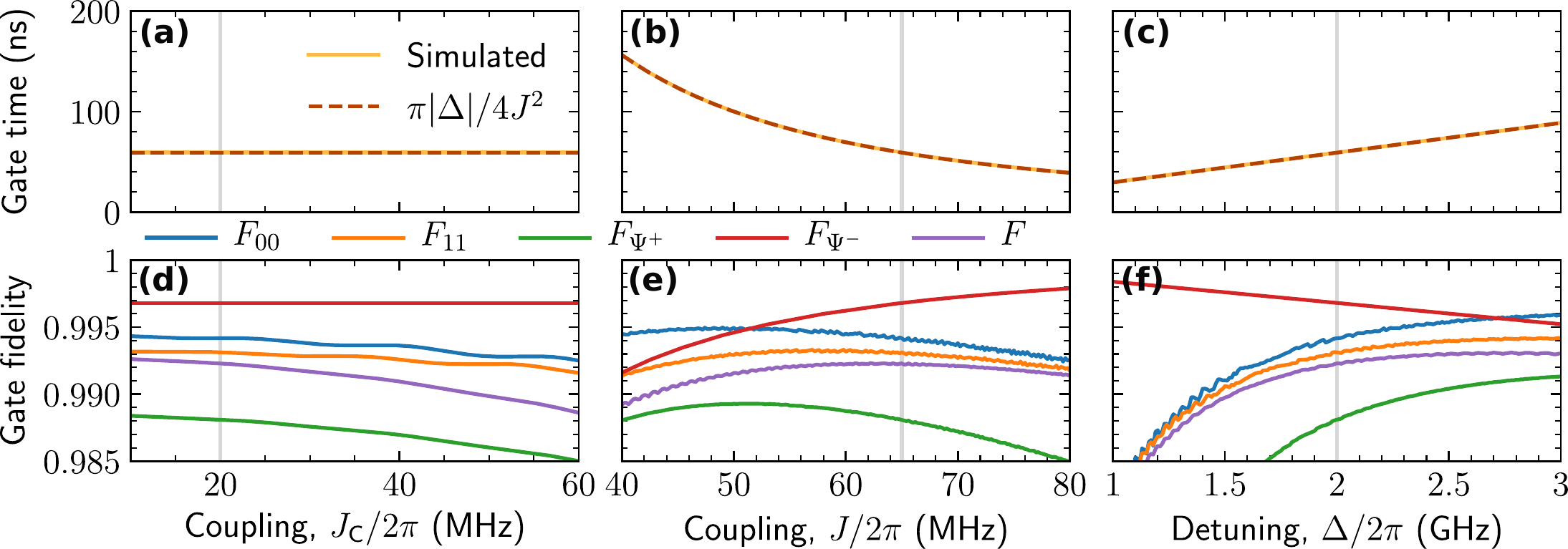}
   \caption{Simulations varying the model parameters $J_\text{C}$, $J$ and $\Delta$, with qubit decoherence of rate $\gamma = \SI{0.01}{\mega\hertz}$. While one parameter is varied, the remaining two are fixed at the values marked by the gray vertical lines (parameter set 1 of Table~\ref{tab:parameters}). \textbf{(a)--(c)} Gate times, also showing the prediction of Eq.~\eqref{eq:tg} as the dashed line. \textbf{(d)--(f)} Gate fidelities, i.e. the fidelities at the simulated gate time.}
   \label{fig:simulation_params}
\end{figure*}

As a quality measure of the gate, we consider the average fidelity\cite{Nielsen2002avfid} (or simply `fidelity' in the following),
\begin{equation}
    \label{eq:av_fidelity_def}
    F(t) \equiv
    \int d\psi \matrixel{\psi}{U_\text{target}^\dagger \mathcal{E}_t(\dyad{\psi}) U_\text{target}}{\psi} \; ,
\end{equation}
which quantifies how well the quantum map $\mathcal{E}_t$ approximates the target unitary gate $U_\text{target}$ over a uniform distribution of input quantum states.
If the diamond gate is run with an arbitrary initial state, the integral is taken over all possible four-qubit states, and can be reduced to a sum over a density matrix basis, as shown in Ref.~\cite{Nielsen2002avfid}.
Putting $U_\text{target} = U$ from Eq.~\eqref{eq:U} and $\mathcal{E}_t(\rho(0)) = \rho(t)$ found from solving Eq.~\eqref{eq:master-equation}, the computed fidelity quantifies the overall performance of the diamond gate with arbitrary initial states.
We denote this fidelity by $F$. Its maximum value (the gate fidelity) defines the gate time, which generally matches the predicted value of Eq.~\eqref{eq:tg} within a few percent.
The sources of gate infidelity are qubit decoherence and state mixing accommodated by a non-zero $J_\text{C}$.

In order to study the performance of the four individual gates of Eqs.~\eqref{eq:UT00}--\eqref{eq:UTpl}, we initialize the control qubits in $\ket{\phi}_\text{C} \in \{ \ket{00}_\text{C}, \ket{11}_\text{C}, \ket{\Psi^+}_\text{C}, \ket{\Psi^-}_\text{C} \}$.
In this case the target operation is a single term in Eq.~\eqref{eq:U}, $U_\text{target} = \dyad{\phi}_\text{C} U_\text{T}^\phi$, and the integral is taken over all states on the form $\ket{\phi}_\text{C}\ket{\psi}_\text{T}$, i.e. only varying the target qubits' state, $\ket{\psi}_\text{T}$.
These states span a subspace of the entire four-qubit Hilbert space characterized by the fixed control state, however couplings to other control states leads to leakage out of the subspace, which we take into account with the appropriate modification of the sum formula in Ref.~\cite{Nielsen2002avfid}. The resulting fidelity is denoted $F_\phi$, and the value at the gate time is denoted the gate fidelity for the associated gate.

Two example parameter sets relevant for superconducting qubits are shown in Table~\ref{tab:parameters}. We use the state-of-the art decoherence rate $\gamma = \SI{0.01}{\mega\hertz}$, corresponding to a qubit life-time of $\gamma^{-1} = \SI{100}{us}$\cite{wang2018}. Figure~\ref{fig:fid_vs_time} shows the simulated fidelities as functions of time. As expected, there is a trade-off between a fast gate and high-fidelity operations.
Parameter set 1 operates in $\SI{59.3}{\nano\second}$ with gate fidelities $\sim 0.99$, which decreases to $\sim 0.96$ for the very fast $\SI{31.5}{\nano\second}$ gate of parameter set 2.
The gate infidelities for each controlled gate follow the expectations discussed in the previous section.
In particular, the idle gate fidelity, $F_{\Psi^-}(t)$, is only limited by qubit decoherence, reducing its value from $1$ to $0.9983$ and $0.9968$, respectively, during the operation time in the two cases.
For the remaining three controlled gates, a longer gate time can improve the gate fidelity, with the drawback of increased susceptibility to qubit decoherence.
Ultimately this limits the number of computations the diamond gate device can run successfully.
For the purpose of demonstrating the model, we will use parameter set 1 in the following, unless otherwise stated.

To probe the sensitivity to the model parameters, we vary each of $\Delta$, $J$ and $J_\text{C}$.
As is evident from Figure~\ref{fig:simulation_params}a--c, the simulated gate times follow closely the prediction of Eq.~\eqref{eq:tg}. Specifically, the gate time is tunable through $\Delta$ and $J$.
The gate fidelities for the individually controlled gates and the total diamond gate are shown in Figure~\ref{fig:simulation_params}a--f. Except for the phase gate controlled by $\ket{\Psi^+}_\text{C}$, which is affected most strongly by couplings to other control states, the fidelities are above 0.99 over a wide range of parameters.
Due to the mathematical equivalence between the two swapping gates controlled by $\ket{00}_\text{C}$ and $\ket{11}_\text{C}$, the gate fidelities for these operations are very similar. We attribute the difference to qubit relaxation, which only affects $\ket{11}_\text{C}$ and becomes more pronounced as the gate time increases.
The identity gate controlled by $\ket{\Psi^-}_\text{C}$ is only limited by decoherence, and its gate fidelity decreases linearly with the gate time.

\begin{figure}
   \includegraphics[width=\columnwidth]{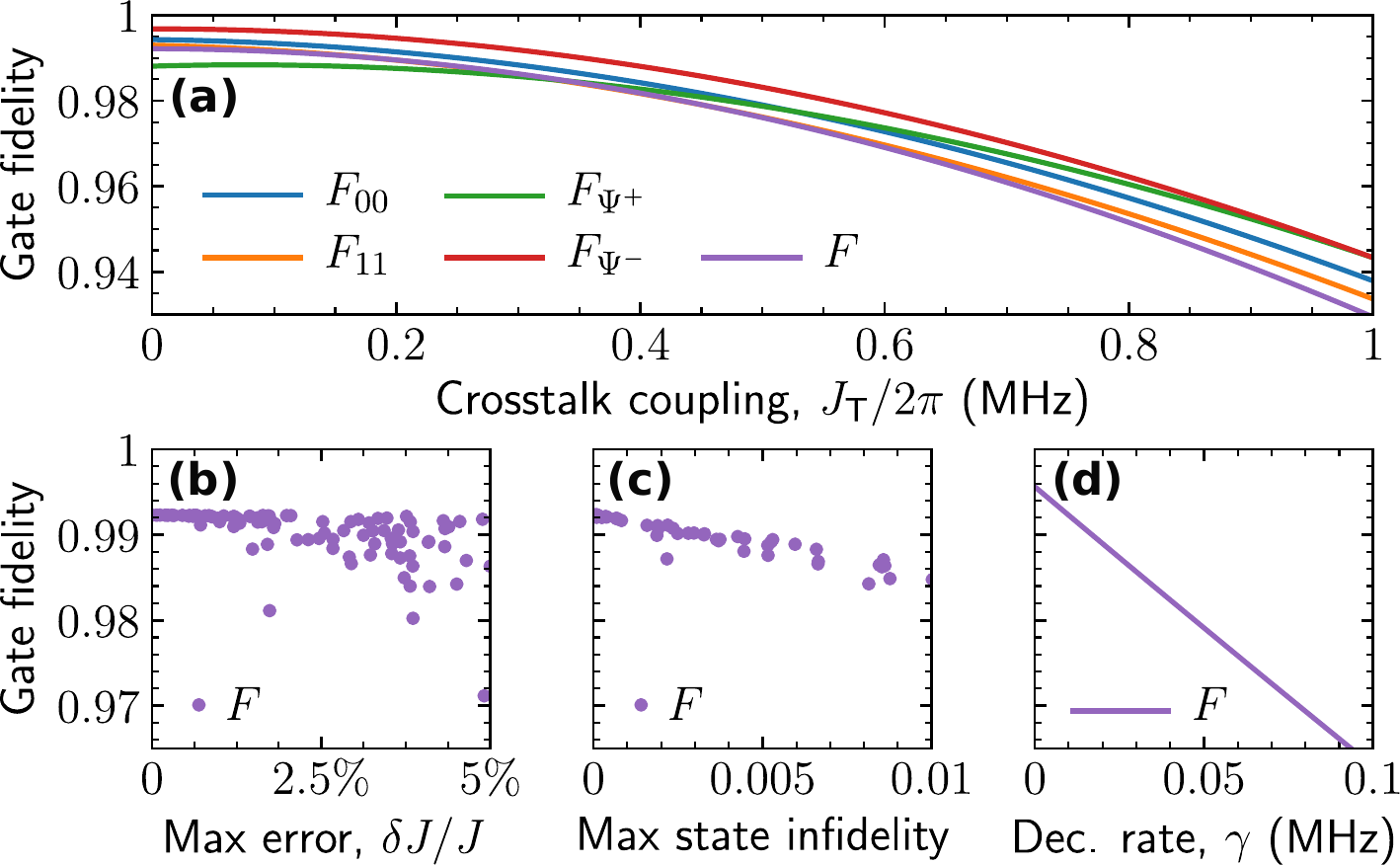}
   \caption{Investigating gate stability for the following system infidelities: \textbf{(a)} Crosstalk coupling between the target qubits. \textbf{(b)} Random asymmetric noise in the couplings between the target and control qubits. \textbf{(c)} Control state infidelity. \textbf{(d)} Qubit decoherence with rate $\gamma$.}
   \label{fig:simulation_noise}
\end{figure}

With a superconducting circuit implementation in mind, we consider a variety of system infidelites and their impact on the gate fidelities, see Figure~\ref{fig:simulation_noise}. Most harmful is a direct capacitive coupling between the target qubits (Figure~\ref{fig:simulation_noise}a), which allows the target qubits to bypass the control qubits, thereby circumventing the interference condition set by the control qubits.
The gate fidelities roughly decrease with the square of the cross-coupling strength $J_\text{T}$, leading to noticable gate infidelities even for a relatively weak coupling. However, as we will show in the next section, crosstalk should not be suppressed, but rather utilized to combat another effect appearing in superconducting qubits: couplings to higher-energy states in the qubits' spectrum.

Figure~\ref{fig:simulation_noise}b shows simulation results with random noise on the couplings between the target and control qubits emulating asymmetries present in an actual circuit due to fabrication limits.
Each data point in the plot corresponds to a simulation with random deviations from the noiseless value, $J$, denoting by $\delta J$ the maximum deviation over the four couplings.
The gate performance is very robust towards this type of noise.

Bell state generation, which is required for the control states $\ket{\Psi^\pm}_\text{C}$, has been shown with a state infidelity of $\sim 0.005$\cite{barends2014cz}.
We introduce control state infidelity in the following way. For each data point in Figure~\ref{fig:simulation_noise}c we contruct a random four-by-four Hermitian matrix $M$, from which we construct a unitary matrix $V = e^{i\epsilon M}$, where $\epsilon$ is a small real parameter.
In the simulations, we apply $V$ to the initial state of the control qubits in order to model imperfect state preparation.
The resulting gate fidelity is shown as a function of the maximum infidelity among the four control states.
The diamond gate suffers a linear decrease in gate fidelity, but remains high-performing for realistic control state infidelity.

Qubit decoherence in the form of relaxation and dephasing is included in the master equation~\eqref{eq:master-equation} with rate $\gamma$.
In Figure~\ref{fig:simulation_noise}d we see that the gate fidelity decreases linearly with $\gamma$.
Even for qubits with $\gamma = \SI{0.05}{\mega\hertz}$, corresponding to a lifetime of $\gamma^{-1} = \SI{20}{\micro\second}$, the gate fidelity is $\sim 0.98$.
We attribute this robustness to the relatively short gate time of $\SI{59.3}{\nano\second}$.

\subsection{Higher-energy states}
\label{sec:qutrits}

In the previous section, we treated a model for four coupled qubits. In the superconducting circuit implementation of Figure~\ref{fig:superconducting_circuit}, these qubits are comprised of the two lowest energy states of the each transmon, $\ket{0}$ and $\ket{1}$.
However, in an actual superconducting circuit, the qubits may couple to higher-energy states in the transmon spectrum, which is the spectrum of a slightly anharmonic oscillator\cite{transmon_original}.
In this section, we analyse the effects from including the second excited state, $\ket{2}$, in the spectrum, thereby turning each qubit into a qutrit.

The full analysis of the circuit of Figure~\ref{fig:superconducting_circuit}b is given in Appendix~\ref{sec:qutrit_analysis}. The resulting four-qutrit Hamiltonian is a sum of the non-interacting part
\begin{equation}
    \label{eq:qutrit_H0}
    \tilde H_0 = -\frac{1}{2} \Omega_\text{T} (\tilde\sigma_z^\text{T1} + \tilde\sigma_z^\text{T2})
                 -\frac{1}{2} \Omega_\text{C} (\tilde\sigma_z^\text{C1} + \tilde\sigma_z^\text{C2}) \; ,
\end{equation}
and the interaction terms
\begin{equation}
    \label{eq:qutrit_Hint}
    \tilde H_\text{int} = J_\text{T} \tilde\sigma_y^\text{T1} \tilde\sigma_y^\text{T2}
                        + J_\text{C} \tilde\sigma_y^\text{C1} \tilde\sigma_y^\text{C2}
                        + J (\tilde\sigma_y^\text{T1} + \tilde\sigma_y^\text{T2}) (\tilde\sigma_y^\text{C1} + \tilde\sigma_y^\text{C2}) \; ,
\end{equation}
which are analogous to Eqs.~\eqref{eq:H0}--\eqref{eq:Hint}. The `Pauli $z$-operator' on qutrit $j$, denoted $\tilde \sigma_j$, includes $\ket{2}_j$ in such a way that it has an energy $\Omega_j + \alpha_j$ above $\ket{1}_j$, with $\Omega_j$ and $\alpha_j$ the frequency and anharmonicity, respectively. Typically $\alpha_j / \Omega_j \sim -0.05$, yielding a small detuning of the second excited state compared to an equidistant spectrum (i.e. to vanishing anharmonicity). The operator is given as
\begin{equation}
    \label{eq:qutrit_sigmaz}
    \tilde\sigma_z^j = \dyad{0}{0}_j - \dyad{1}{1}_j - \left( 3 + \frac{2\alpha_j}{\Omega_j} \right) \dyad{2}{2}_j \; ,
\end{equation}
The `Pauli $y$-operator' on qutrit $j$ is
\begin{equation}
    \label{eq:qutrit_sigmay}
    \tilde\sigma_y^j = i T_0^j \dyad{1}{0}_j + i T_2^j \dyad{2}{1}_j + \text{H.c.} \; ,
\end{equation}
where $T_0^j \approx 1$ and $T_2^j \approx \sqrt 2$ can be expressed in terms of $\Omega_j$ and $\alpha_j$ (see Appendix~\ref{sec:qutrit_analysis}).
Hence, the coupling between the first and second excited state is as strong as the coupling between the two lowest (qubit) levels.
Due to the small anharmonicity in transmons, i.e. that the energy separation between the qubit levels almost equals the separation between the first and second excited state, couplings that exchange a single excitation like $\ket{11} \rightarrow \ket{02}$ are not strongly energetically suppressed.
In fact, this transition is sometimes used for the \textsf{CZ} gate\cite{krantz2019}.
Notice that this lack of suppression holds for transmons in general, and is not a consequence of the specific model considered here.

This has two undesired consequences. Firstly, unless $\abs{J_\text{C}/\alpha_\text{C}} \ll 1$, it allows the control state $\ket{11}_\text{C}$ to mix with $\ket{02}_\text{C}$ and $\ket{20}_\text{C}$, leading to a non-conserved control state during the gate operation. This can be resolved by redefining the control state as
\begin{equation}
    \label{eq:tilde11}
    \ket{\tilde{11}}_\text{C} = \cos\tilde\theta \ket{11}_\text{C} + \sin\tilde\theta \frac{1}{\sqrt 2} (\ket{02}_\text{C} + \ket{20}_\text{C}) \; ,
\end{equation}
with the mixing angle $\tilde\theta = -\frac{1}{2} \arctan (2\sqrt 2 J_\text{C} T_1^\text{C} T_2^\text{C} / \alpha_\text{C}) \sim 0.5$, such that it is an eigenstate of an effective control state Hamiltonian.
This introduces a significant component of $(\ket{02}_\text{C} + \ket{20}_\text{C})/\sqrt{2}$, which is avoided if $J_\text{C} = 0$.
Details are found in Appendix~\ref{sec:qutrit_analysis}.

\begin{figure}
   \includegraphics[width=\columnwidth]{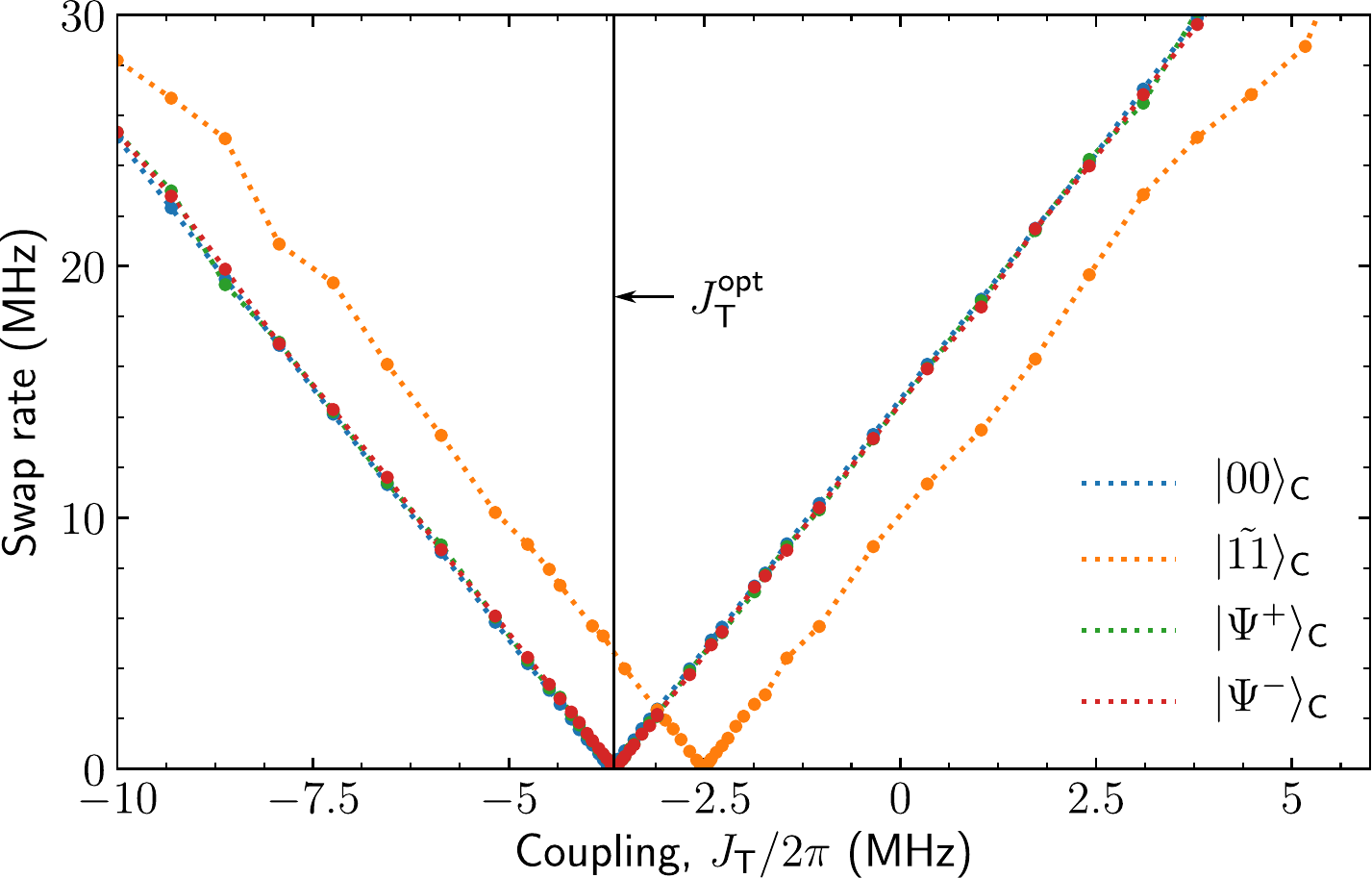}
   \caption{Swap rate, found as the inverse of the smallest time $t$ where the swap fidelity (probability) $\vert \bra{\phi}_\text{C}\bra{01}_\text{T} e^{-i (\tilde H_0 + \tilde H_\text{int})t} \ket{10}_\text{T} \ket{\phi}_\text{C} \vert^2$ becomes close to unity, versus crosstalk strength $J_\text{T}$.
   Data points are shown with the control state $\ket{\phi}_\text{C}$ set to each of the displayed states.
   The parameters used in the simulation are
   $J_\text{C}/2\pi = \SI{20}{\mega\hertz}$, $J/2\pi = \SI{65}{\mega\hertz}$, $\Omega_\text{C}/2\pi = \SI{7}{\giga\hertz}$,
   $\Omega_\text{T}/2\pi = \SI{9}{\giga\hertz}$,
   $\alpha_\text{C} = \SI{-270}{\mega\hertz}$ and $\alpha_\text{T} = \SI{-280}{\mega\hertz}$.
   The optimal value of Eq.~\eqref{eq:JTopt} is marked with a vertical line, $J_\text{T}^\text{opt}/2\pi = \SI{-3.66}{\mega\hertz}$.}
   \label{fig:qutrit_swap_time_vs_JT}
\end{figure}

\begin{figure*}
   \includegraphics[width=2\columnwidth]{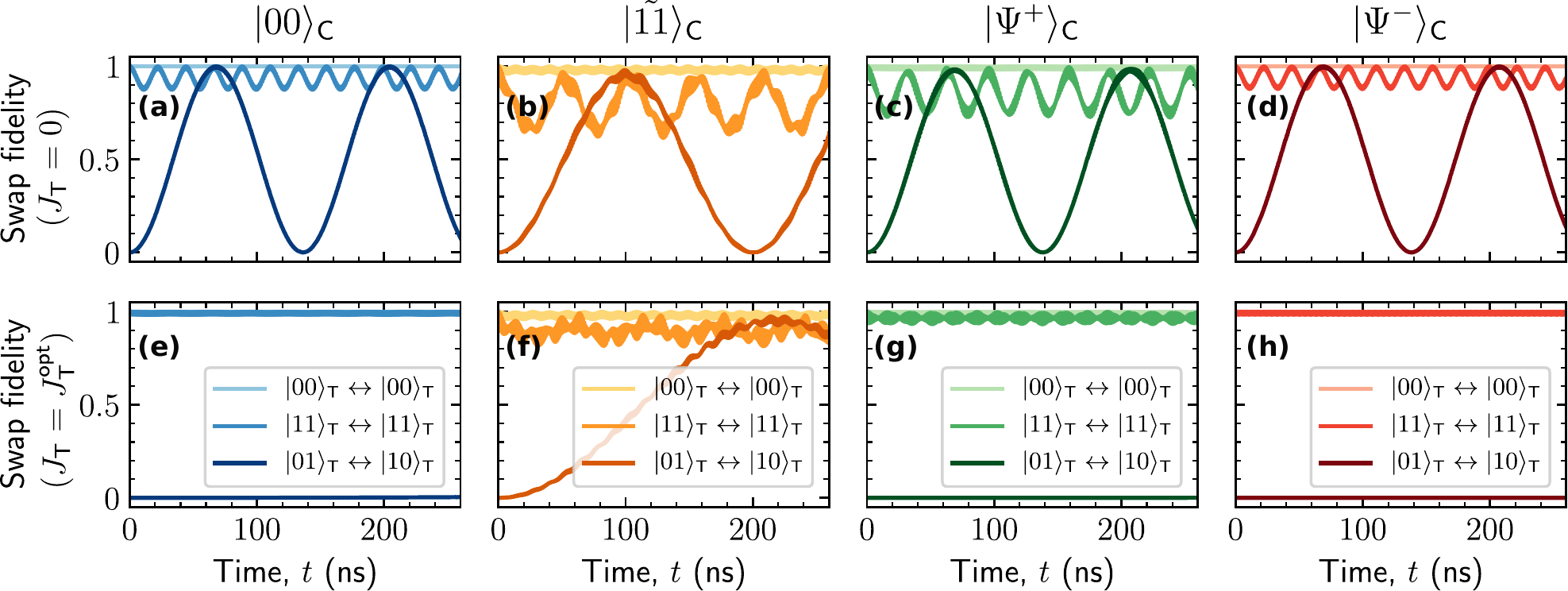}
   \caption{Fidelity for swapping $\ket{\psi}_\text{T} \leftrightarrow \ket{\psi'}_\text{T}$ for the indicated processes, computed as
   $\vert \bra{\phi}_\text{C}\bra{\psi'}_\text{T} e^{-i (\tilde H_0 + \tilde H_\text{int})t} \ket{\psi}_\text{T} \ket{\phi}_\text{C} \vert^2$, with the control state $\ket{\phi}_\text{C}$ indicated above each column.
   The parameters used in the simulation are
   $J_\text{C}/2\pi = \SI{20}{\mega\hertz}$, $J/2\pi = \SI{65}{\mega\hertz}$, $\Omega_\text{C}/2\pi = \SI{7}{\giga\hertz}$,
   $\Omega_\text{T}/2\pi = \SI{9}{\giga\hertz}$,
   $\alpha_\text{C} = \SI{-270}{\mega\hertz}$ and $\alpha_\text{T} = \SI{-280}{\mega\hertz}$.
   \textbf{(a)--(d)} No crosstalk, $J_\text{T} = 0$. \textbf{(e)--(h)} Crosstalk is set to its optimal value of Eq.~\eqref{eq:JTopt}, $J_\text{T}^\text{opt}/2\pi = \SI{-3.66}{\mega\hertz}$.}
   \label{fig:qutrit_swap_fids_vs_time}
\end{figure*}

Secondly, excitations to the second excited states allow unwanted processes which bypass the control.
For instance, when the diamond gate is desired to be idle, leakage across the control can occur via:
\begin{equation}
    \label{eq:qutrit_leakage}
    \ket{\Psi^-}_\text{C} \ket{10}_\text{T}
    \rightarrow
    \frac{1}{\sqrt 2} ( \ket{02}_\text{C} - \ket{20}_\text{C} ) \ket{00}_\text{T}
    \rightarrow
    \ket{\Psi^-}_\text{C} \ket{01}_\text{T} \; .
\end{equation}
Since this is a second order process in the qutrit model Hamiltonian, it would not pose a threat to the functionality of the diamond gate if it only relied on (generally faster) first order processes. However, the swap operations of Eqs.~\eqref{eq:UT00}--\eqref{eq:UT11} are also second order processes, leading to a failure of the idle diamond gate on the same time-scale as the operation of the swap gates. Similarly, the control state $\ket{\Psi^+}$ fails to prevent excitation leakage across the control, corrupting the operation of  Eq.~\eqref{eq:UTpl}.

However, these undesired processes can be mitigated by taking advantage of the effects of crosstalk. The circuit analysis in Appendix~\ref{sec:qutrit_analysis} reveals a weak unavoidable crosstalk coupling of strength $J_\text{T}$ in the interaction Hamiltonian~\eqref{eq:qutrit_Hint}, which by itself has a significant negative impact on the gate fidelities, c.f. Figure~\ref{fig:simulation_noise}a.
This leads directly to leakage across the control through processes of the type
\begin{equation}
    \label{eq:crosstalk_leakage}
    \ket{\Psi^-}_\text{C} \ket{10}_\text{T}
    \rightarrow
    \ket{\Psi^-}_\text{C} \ket{01}_\text{T} \; .
\end{equation}
This process has the same unwanted outcome as the one of Eq.~\ref{eq:qutrit_leakage}. As we show below, we can therefore restore the gate functionality by tuning the value of $J_\text{T}$ such that these two unwanted leakage processes cancel each other.
Analyzing the problem with second order perturbation theory in order to calculate the amplitude of the leaked state (see Appendix~\ref{sec:qutrit_analysis}), we find destructive interference between these processes when the crosstalk strength takes the optimal value
\begin{equation}
    \label{eq:JTopt}
    \begin{aligned}
        J_\text{T}^\text{opt} ={}
        &\frac{(J T_2^\text{C})^2}{\Omega_\text{C} + \Omega_\text{T} + \alpha_\text{C} + J_\text{C}(T_1^\text{C})^2} \\
        &+ \frac{(J T_2^\text{C})^2}{\Omega_\text{C} - \Omega_\text{T} + \alpha_\text{C} + J_\text{C}(T_1^\text{C})^2} \; .
    \end{aligned}
\end{equation}
Thus by tuning the crosstalk strength to $J_\text{T} = J_\text{T}^\text{opt}$, we expect the fidelity for the target qubit swap $\ket{01}_\text{T} \leftrightarrow \ket{10}_\text{T}$ to diminish, or equivalently a vanishing swap rate, when the control state is $\ket{\Psi^\pm}_\text{C}$.
Figure~\ref{fig:qutrit_swap_time_vs_JT} shows the swap rate for varying $J_\text{T}$, with control qubits in each of the four control states.
We find two distinct zero-points, one for the data related to the control states $\ket{00}_\text{C}$ and $\ket{\Psi^\pm}_\text{C}$ at the expected value $J_\text{T}^\text{opt}$ (vertical line), and one for $\ket{\tilde{11}}_\text{C}$.
Thus, it is possible to prevent the unwanted swap operation for the control states $\ket{\Psi^\pm}_\text{C}$, but as a consequence also the swap operation controlled by $\ket{00}_\text{C}$ is obstructed.
On the other hand, the swap operation controlled by $\ket{\tilde{11}}_\text{C}$ is preserved at $J_\text{T} = J_\text{T}^\text{opt}$, although the gate time is prolonged to around $\SI{220}{\nano\second}$.
Remarkably, for $J_\text{T}/2\pi \approx \SI{-2.5}{\mega\hertz}$ the situation is reversed. Here, putting the control in $\ket{\tilde{11}}_\text{C}$ prevents swapping, while the three remaining control states permit it.
At each zero-point, the gate time (inverse swap rate) for the swapping gate(s) is prolonged compared to the results in the previous section.
To reduce the gate time, one should pick parameters such that the zero-points are further apart, or such that the inclination of the graphs are steeper.

Figure~\ref{fig:qutrit_swap_fids_vs_time} illustrates in more detail the cancellation of unwanted transfer by crosstalk engineering.
Each subfigure shows the swap fidelity for different initial target qubit states. The control is initialized in the state indicated above each column.
Figure~\ref{fig:qutrit_swap_fids_vs_time}a--d (the top row) show simulations for $J_\text{T} = 0$, while the crosstalk has been put to its optimal value, $J_\text{T} = J_\text{T}^\text{opt}$, in Figure~\ref{fig:qutrit_swap_fids_vs_time}e--h (the bottom row).
As expected from Figure~\ref{fig:qutrit_swap_time_vs_JT}, the swap $\ket{01}_\text{T} \leftrightarrow \ket{10}_\text{T}$ (dark lines) occurs for any control state when there is no crosstalk, but is controlled uniquely by $\ket{\tilde{11}}_\text{C}$ when the crosstalk is at the optimal value.
In the cases of $\ket{00}_\text{T}$ and $\ket{11}_\text{T}$, we wish to maintain a unit fidelity across all control states, i.e. the states should acquire at most a phase.
Tuning the crosstalk to $J_\text{T}^\text{opt}$ also improves the gate operation in this regard.

Engineering crosstalk to mitigate unwanted leakage through higher-excited states is killing two birds with one stone: Each process is harmful to the functionality of the diamond gate, but letting them cancel each other preserves the ability to control the swap operation.
The price is the loss of swap functionality in the gate controlled by $\ket{00}_\text{C}$, and an increased gate time for the model parameters considered here.
Generally, the phases applied to each target state will be modified for all four controlled gates, but we do not pursue an analysis here, as other factors specific to the implementation will contribute to this as well.
Rather, our main goal was to demonstrate a passive method for dealing with undesired leakage processes.

\section{Discussion}

We have proposed a quantum interference device by coupling four qubits with exchange interactions.
By analyzing the unitary dynamics of the system, we have shown that it realizes the diamond gate: a four-way controlled two-qubit gate, with the ability to run two different entangling swap and phase operations, a (parity) phase operation, an idling gate with no dynamics, or an arbitrary superposition of these.
We considered an implementation in superconducting qubits using transmon qubits, and found that it generally operated fast and with high fidelity using state-of-the-art model and noise parameters. When taking second excited states into account, we had to prevent leakage across the control by engineering crosstalk, demonstrating a general method to avoid leakage in superconducting qubit systems.
The cost of this was a single redefined control state, one swap gate turning into a phase gate, altered phases on the gates, and a slower gate for the considered parameters.
However, we only consider this analysis a starting point for an actual implementation, which might also include active microwave driving to optimize the operations or to prevent certain transitions.
It might also be worthwile to consider other types of superconducting qubits with larger anharmonicity, or entirely different platforms such as lattices of ultracold atoms or ions, where qubit encoded in hyperfine states or vibrational modes are far detuned from the rest of the spectrum.

We illustrated how the four-qubit diamond gate device can constitute an essential building block in an extensible quantum computer, and proposed a simple scheme where quantum algorithms are run on the computer by parallel processing on each four-qubit module interspersed with two-qubit operations spreading entanglement in the system, and single-qubit operations.
Evidently, this scheme is adaptable to many different algorithms, and future work will investigate which algorithms are suitable to be implemented in the diamond-plaquette device.

\begin{acknowledgments}
  This research was funded in part by the U.S. Army Research Office Grant No. W911NF-17-S-0008. N.J.S.L., L.B.K., and N.T.Z. acknowledge support from the Carlsberg Foundation and The Danish National Research Council under the Sapere Aude program. M.K. acknowledges support from the Carlsberg Foundation. T.W.L. acknowledges support from Microsoft. N.J.S.L. ackowledge discussions with Alán Aspuru-Guzik, Daniel Kyungdeock Park, Kasper Sangild, and Stig Elkjær Rasmussen. The views and conclusions contained herein are those of the authors and should not be interpreted as necessarily representing the official policies or endorsements of the US Government.
\end{acknowledgments}

\onecolumngrid
\appendix

\section{Unitary dynamics in the qubit model}
\label{sec:unitary_gate_analysis}

In this appendix we show that the Hamiltonian of Eq.~\eqref{eq:H} realizes the four-qubit quantum gate of Eq.~\eqref{eq:U} by analyzing the dynamics within Floquet theory. Typically in superconducting qubits $\lvert \Delta \rvert \gg \lvert J \rvert, \lvert J_\text{C} \rvert$, so if we think of the qubit detuning, $\Delta$, as a driving frequency, the system is driven rapidly compared to the time-scale set by the qubit interaction strengths. Consequently, on the gate operation time-scale, it is appropriate to consider the Magnus expansion for the Floquet Hamiltonian to first order in $J/\Delta$, which can be computed as\cite{polkovnikov2015}:
\begin{equation}
\begin{aligned}
  H_\text{F} ={} &J_\text{C} \, (\sigma_+^\text{C1}\sigma_-^\text{C2} + \sigma_-^\text{C1}\sigma_+^\text{C2})
          + \frac{J^2}{\Delta} (\sigma_-^\text{T1} + \sigma_-^\text{T2})(\sigma_+^\text{T1} + \sigma_+^\text{T2})(\sigma_z^\text{C1} + \sigma_z^\text{C2})
          - \frac{J^2}{\Delta} (\sigma_-^\text{C1} + \sigma_-^\text{C2})(\sigma_+^\text{C1} + \sigma_+^\text{C2})(\sigma_z^\text{T1} + \sigma_z^\text{T2}) \\
          &- \frac{J_\text{C}J}{\Delta} (\sigma_+^\text{C1}\sigma_z^\text{C2} + \sigma_+^\text{C2}\sigma_z^\text{C1})(\sigma_-^\text{T1} + \sigma_-^\text{T2})
          - \frac{J_\text{C}J}{\Delta} (\sigma_-^\text{C1}\sigma_z^\text{C2} + \sigma_-^\text{C2}\sigma_z^\text{C1})(\sigma_+^\text{T1} + \sigma_+^\text{T2}) \; .
\end{aligned}
\label{eq:HF}
\end{equation}
Within the Floquet formalism $\exp(-iH_\text{F}T)$ takes the system from time zero through one driving cycle of period $T=2\pi/\lvert\Delta\rvert$. Successive application $n$ times yields the time-evolution operator, $U(nT) = \exp(-iH_\text{F} nT)$. Since the gate time is much larger than one period, we consider $t = nT$ a continuous time variable, and the continuous time-evolution operator, $U(t) = \exp(-iH_\text{F} t)$.

Suppose we initialized the control qubits in one of the control basis states, $\{ \ket{00}_\text{C}, \ket{11}_\text{C}, \ket{\Psi^+}_\text{C}, \ket{\Psi^-}_\text{C} \}$. Typically, one thinks of control qubits, or their state, as a catalyzer for a given gate operation performed on the target qubits. The control qubits are allowed to partake in the gate operation, for instance by facilitating state transfer between target qubits not directly coupled, as long as the control qubits return to their initial state after the completion of the gate operation. A priori we cannot guarantee that this is the case. In fact, we see by application of the Floquet Hamiltonian $H_\text{F}$ of Eq.~\eqref{eq:HF} to each control state (producing operators acting on the target qubits only) that they generally evolve in time:
\begin{align}
    \label{eq:HF_on_00}
    &H_\text{F} \ket{00}_\text{C} = \ket{00}_\text{C} \frac{2J^2}{\Delta}
    \left[ (\sigma_-^\text{T1} + \sigma_-^\text{T2})(\sigma_+^\text{T1} + \sigma_+^\text{T2}) - \sigma_z^\text{T1} - \sigma_z^\text{T2} \right]
    - \ket{\Psi^+}_\text{C} \frac{\sqrt{2} J_\text{C} J}{\Delta} (\sigma_-^\text{T1} + \sigma_-^\text{T2}) \; , \\
    \label{eq:HF_on_11}
    &H_\text{F} \ket{11}_\text{C} = - \ket{11}_\text{C} \frac{2J^2}{\Delta}
    (\sigma_-^\text{T1} + \sigma_-^\text{T2})(\sigma_+^\text{T1} + \sigma_+^\text{T2})
    + \ket{\Psi^+}_\text{C} \frac{\sqrt{2} J_\text{C} J}{\Delta} (\sigma_+^\text{T1} + \sigma_+^\text{T2}) \; , \\
    \label{eq:HF_on_Psi+}
    &H_\text{F} \ket{\Psi^+}_\text{C} = \ket{\Psi^+}_\text{C} \left[ J_\text{C} - \frac{2J^2}{\Delta} (\sigma_z^\text{T1} + \sigma_z^\text{T2}) \right]
    + \ket{11}_\text{C} \frac{\sqrt{2} J_\text{C}J}{\Delta} (\sigma_-^\text{T1} + \sigma_-^\text{T2})
    - \ket{00}_\text{C} \frac{\sqrt{2} J_\text{C}J}{\Delta} (\sigma_+^\text{T1} + \sigma_+^\text{T2}) \; , \\
    \label{eq:HF_on_Psi-}
    &H_\text{F} \ket{\Psi^-}_\text{C} = \ket{\Psi^-}_\text{C} (-J_\text{C}) \; .
\end{align}
We see that $H_\text{F}$ couples the triplet states $\ket{00}_\text{C}$, $\ket{11}_\text{C}$ and $\ket{\Psi^+}_\text{C}$, but that the singlet state $\ket{\Psi^-}_\text{C}$ is unchanged in time. Notice that all control states decouples in the special case $J_\text{C} = 0$, i.e. when there is no direct coupling between the control qubits.

\subsection{The case of $J_\text{C} = 0$}

In this case, each control state is perfectly preserved under the time-evolution, and we can simply determine the gate operation on the target qubits associated with each control state. However, the absence of a direct coupling between the control qubits makes it difficult to prepare the entangled Bell states, $\ket{\Psi^\pm}_\text{C}$.
Ideally, the control-control coupling would be tunable and only on during control state preparation.
On the other hand, since it does not couple to any of the target qubits, we do not expect the value of $J_\text{C}$ to be of fundamental importance to the nature of the gate operations, which is our main focus here. Assuming $J_\text{C} = 0$, the Floquet Hamiltonian can be cast as
\begin{equation}
    H_\text{F} = \dyad{00}_\text{C} H^{00}_\text{T} + \dyad{11}_\text{C} H^{11}_\text{T} + \dyad{\Psi^+}_\text{C} H^{\Psi^+}_\text{T} + \dyad{\Psi^-}_\text{C} H^{\Psi^-}_\text{T} \; ,
\end{equation}
with the following Hamiltonians acting only on the target qubits:
\begin{align}
    \label{eq:H00}
    &H^{00}_\text{T} = \frac{2J^2}{\Delta} \left[ (\sigma_-^\text{T1} + \sigma_-^\text{T2})(\sigma_+^\text{T1} + \sigma_+^\text{T2}) - \sigma_z^\text{T1} - \sigma_z^\text{T2} \right] \; , \\
    \label{eq:H11}
    &H^{11}_\text{T} = - \frac{2J^2}{\Delta} (\sigma_-^\text{T1} + \sigma_-^\text{T2})(\sigma_+^\text{T1} + \sigma_+^\text{T2}) \; , \\
    \label{eq:HPsi+}
    &H^{\Psi^+}_\text{T} = - \frac{2J^2}{\Delta} (\sigma_z^\text{T1} + \sigma_z^\text{T2}) \; , \\
    \label{eq:HPsi-}
    &H^{\Psi^-}_\text{T} = 0 \; .
\end{align}
In order to compute the time-evolution operator, $U(t) = \exp(-i H_\text{F} t)$, we notice that $H_\text{F}$ is on the form
\begin{equation}
    H_\text{F} = \sum_{i=1}^N P_i H_i \; ,
\end{equation}
where $P_i = \dyad{i}$ is the projector onto the $i$'th orthonormal basis state of the $N$-dimensional subsystem $A$, and $H_i$ is a Hamiltonian on a disjoint subsystem $B$, such that $H_i$ commute with every $P_j$. Operators on this form has the property that the product of any two terms is zero, $(P_i H_i)(P_j H_j) = 0$ for $i \neq j$, enabling an algebraic property known as ``freshman's dream'': $(H_\text{F})^n = \sum_{i=1}^N (P_i H_i)^n$ for any integer $n > 0$. This has the consequence that the operator exponential can be written as a sum:
\begin{equation}
    \exp(-i H_\text{F} t) = \sum_{n=0}^\infty \frac{(-it)^n}{n!} (H_\text{F})^n 
    = 1 + \sum_{n=1}^\infty \frac{(-it)^n}{n!} \sum_{i=1}^N (P_i H_i)^n
    = 1 - N + \sum_{i=1}^N \exp(-i P_i H_i t)\; .
\end{equation}
Since $(P_i)^n = P_i$ for any integer $n>0$, we can pull the projector out of each exponential in the sum:
\begin{equation}
    \exp(-i P_i H_i t) = \sum_{n=0}^\infty \frac{(-it)^n}{n!} (P_i H_i)^n
    = 1 + \sum_{n=1}^\infty \frac{(-it)^n}{n!} P_i (H_i)^n
    = 1 - P_i + P_i \exp(-i t H_i) \; .
\end{equation}
Finally, utilizing $\sum_{i=1}^N P_i = 1$, we find that the time-evolution operator can be expressed as
\begin{equation}
    U(t) = \exp(-i H_\text{F} t)
    = 1 - N + \sum_{i=1}^N \left[ 1 - P_i + P_i \exp(-i t H_i) \right]
    = \sum_{i=1}^N P_i \exp(-i t H_i) \; .
\end{equation}
The above decomposition of the time-evolution can used whenever one or more control qubits (subsystem $A$) catalyze a unitary gate operation on a set of target qubits (subsystem $B$) in the sense that the Hamiltonian does not mix the chosen control states. In our case, we can easily express the Hamiltonians~\eqref{eq:H00}--\eqref{eq:HPsi-} as matrices and find the unitary matrix exponentials. In the computational basis of the target qubits, they are as follows:
\begin{align}
    \label{eq:U00}
    &U_\text{T}^{00}(t) = \exp(-i H_\text{T}^{00} t)
    =
    \begin{pmatrix}
        \makebox[2em]{$1$} & 0 & 0 & 0 \\[.2em]
        0 & \makebox[5em]{$\frac{1}{2}e^{-it\zeta} + \frac{1}{2}$} & \makebox[5em]{$\frac{1}{2}e^{-it\zeta} - \frac{1}{2}$} & 0 \\[.2em]
        0 & \makebox[5em]{$\frac{1}{2}e^{-it\zeta} - \frac{1}{2}$} & \makebox[5em]{$\frac{1}{2}e^{-it\zeta} + \frac{1}{2}$} & 0 \\[.2em]
        0 & 0 & 0 & \makebox[2em]{$e^{-it\zeta}$}
    \end{pmatrix} \; , \\
    \label{eq:U11}
    &U_\text{T}^{11}(t) = \exp(-i H_\text{T}^{11} t)
    =
    \begin{pmatrix}
        \makebox[2em]{$e^{it\zeta}$} & 0 & 0 & 0 \\[.2em]
        0 & \makebox[5em]{$\frac{1}{2}e^{it\zeta} + \frac{1}{2}$} & \makebox[5em]{$\frac{1}{2}e^{it\zeta} - \frac{1}{2}$} & 0 \\[.2em]
        0 & \makebox[5em]{$\frac{1}{2}e^{it\zeta} - \frac{1}{2}$} & \makebox[5em]{$\frac{1}{2}e^{it\zeta} + \frac{1}{2}$} & 0 \\[.2em]
        0 & 0 & 0 & \makebox[2em]{$1$}
    \end{pmatrix} \; , \\
    \label{eq:UPsi+}
    &U_\text{T}^{\Psi^+}(t) = \exp(-i H_\text{T}^{\Psi^+} t)
    =
    \begin{pmatrix}
        \makebox[2em]{$e^{it\zeta}$} & 0 & 0 & 0 \\[.2em]
        0 & \makebox[2em]{$1$} & 0 & 0 \\[.2em]
        0 & 0 & \makebox[2em]{$1$} & 0 \\[.2em]
        0 & 0 & 0 & \makebox[2em]{$e^{-it\zeta}$}
    \end{pmatrix} \; , \\
    \label{eq:UPsi-}
    &U_\text{T}^{\Psi^-}(t) = \exp(-i H_\text{T}^{\Psi^-} t)
    =
    \begin{pmatrix}
        \makebox[2em]{$1$} & 0 & 0 & 0 \\[.2em]
        0 & \makebox[2em]{$1$} & 0 & 0 \\[.2em]
        0 & 0 & \makebox[2em]{$1$} & 0 \\[.2em]
        0 & 0 & 0 & \makebox[2em]{$1$}
    \end{pmatrix} \; ,
\end{align}
with $\zeta = 4J^2 / \Delta$. The time-evolution operator for the four-qubit system is then
\begin{equation}
    \label{eq:UofT_noJC}
    U(t) = \dyad{00}_\text{C} U_\text{T}^{00}(t) + \dyad{11}_\text{C} U_\text{T}^{11}(t)
         + \dyad{\Psi^+}_\text{C} U_\text{T}^{\Psi^+}(t) + \dyad{\Psi^-}_\text{C} U_\text{T}^{\Psi^-}(t) \; .
\end{equation}
Thus, each of the four unitaries~\eqref{eq:U00}--\eqref{eq:UPsi-} above is a gate operation performed on the target qubits, controlled entirely by the four control states, which are unaltered by the operation. The control states $\ket{00}_\text{C}$ and $\ket{11}_\text{C}$ induce oscillations between the target qubit states combined with a phase on either $\ket{00}_\text{T}$ or $\ket{11}_\text{T}$, depending on the control state, and $\ket{\Psi^+}_\text{C}$ controls a pure phase operation that distinguishes between the number of excitations in the target qubits.
The singlet control state, $\ket{\Psi^-}_\text{C}$, on the other hand, does nothing to the target qubits, and this control state can therefore be used to turn off the gate between the target qubits. The gate is fully quantum mechanical, as superpositions of control states will run the corresponding computations on the target qubits in parallel. The system comprise a true four-qubit quantum interference device in the form of a four-way controlled two-qubit gate (the diamond gate).

Of particular interest is the gate operation at the time $t = t_g \equiv \pi / \lvert \zeta \rvert$, which results in the operations discussed in the main text. Setting $t=t_g$ in Eq.~\eqref{eq:UofT_noJC} produces the four-qubit unitary gate $U$ of Eq.~\eqref{eq:U}.


As we shall see below, a non-zero $J_\text{C}$ introduce infidelities, albeit only very small.
We may therefore ask ourselves whether the control-qubit coupling is necessary at all. After all, if the control qubits are decoupled, $J_\text{C}=0$, we may still initialize them in the computational basis states by driving each control qubit individually. Expressing the Bell states in the computational basis casts Eq.~\eqref{eq:UofT_noJC} as:
\begin{equation}
    \begin{aligned}
    U(t) ={} &\dyad{00}_\text{C} U_\text{T}^{00}(t) + \dyad{11}_\text{C} U_\text{T}^{11}(t)
    + (\dyad{01}_\text{C} + \dyad{10}_\text{C}) \frac{1}{2} ( U_\text{T}^{\Psi^+}(t) + U_\text{T}^{\Psi^-}(t)) \\
    &+ (\dyad{01}{10}_\text{C} + \dyad{10}{01}_\text{C}) \frac{1}{2} ( U_\text{T}^{\Psi^+}(t) - U_\text{T}^{\Psi^-}(t)) \; .
    \end{aligned}
\end{equation}
Notice that the computational basis control states are unaltered if and only if $U_\text{T}^{\Psi^+}(t) = U_\text{T}^{\Psi^-}(t)$, or equivalently $t = 0, 2t_g, 4t_g, \dots$, which reduces $U(t)$ to the identity operator on all qubits. This is not surprising, since $J_\text{C} = 0$ and our choice of bases results in complete symmetry between the control and target qubits, and if we require no evolution of the control qubits, no evolution of the target qubits can occur either. On the other hand, the symmetry betwen control and target qubits when $J_\text{C}=0$ means that the role of control and target qubits is only a matter of choice of basis. This also means that the roles can be interchanged between operations, for instance in a larger quantum computer where the four-qubit diamond gate device is a subsystem.

\subsection{The case of non-zero $J_\text{C}$}

When $J_\text{C}$ is non-zero, we see from Eqs.~\eqref{eq:HF_on_00}--\eqref{eq:HF_on_Psi+} that the Floquet Hamiltonian couples the triplet control states  $\{\ket{00}_\text{C}, \ket{11}_\text{C}, \ket{\Psi^+}_\text{C} \}$. In the following we study how strongly they mix during the gate operation, and we find that it only has a weak impact on the gate fidelity.

Before we proceed with the calculation, we notice from the first terms on the right-hand side of Eqs.~\eqref{eq:HF_on_Psi+}--\eqref{eq:HF_on_Psi-} that the presence of $J_\text{C}$ adds a global phase to the pure phase gates.
Specifically, the gates of Eqs.~\eqref{eq:UPsi+}--\eqref{eq:UPsi-} must be modifies $U_\text{T}^{\Psi^\pm}(t) \rightarrow e^{\mp it J_\text{C}} U_\text{T}^{\Psi^\pm}(t)$, leading to the expressions \eqref{eq:UTpl}--\eqref{eq:UTmi} in the main text at $t=t_g$.
As the singlet state $\ket{\Psi^-}_\text{C}$ remains uncoupled to other control states, this is the only modification of the identity gate, which thus suffers no infidelity due to the control qubit coupling.

Since $\ket{00}_\text{C}$ and $\ket{11}_\text{C}$ couples to $\ket{\Psi^+}_\text{C}$ in a completely analogous way, it is enough to treat the case $\ket{11}_\text{C}$.
Suppose we initialize the control qubits in $\ket{11}_\text{C}$, and consider the effect of $H_\text{F}$ on each four-qubit state, expressing the target-qubit states in the basis $\{\ket{00}_\text{T}, \ket{11}_\text{T}, \ket{\Psi^+}_\text{T}, \ket{\Psi^-}_\text{T} \}$:
\begin{align}
    &H_\text{F} \, \ket{11}_\text{C} \ket{00}_\text{T}
    = \frac{2J_\text{C}J}{\Delta} \ket{\Psi^+}_\text{C} \ket{\Psi^+}_\text{T} - \frac{4J^2}{\Delta} \ket{11}_\text{C} \ket{00}_\text{T} \; ,
    \label{eq:HF_on_1100} \\
    &H_\text{F} \, \ket{11}_\text{C} \ket{11}_\text{T} = 0 \; ,
    \label{eq:HF_on_1111} \\
    &H_\text{F} \, \ket{11}_\text{C} \ket{\Psi^+}_\text{T}
    = \frac{2J_\text{C}J}{\Delta} \ket{\Psi^+}_\text{C} \ket{11}_\text{T} - \frac{4J^2}{\Delta} \ket{11}_\text{C} \ket{\Psi^+}_\text{T} \; ,
    \label{eq:HF_on_11Psi+} \\
    &H_\text{F} \, \ket{11}_\text{C} \ket{\Psi^-}_\text{T} = 0 \; .
    \label{eq:HF_on_11Psi-}
\end{align}
Starting with Eq.~\eqref{eq:HF_on_11Psi+}, we see that $H_\text{F}$ couples $\ket{11}_\text{C}\ket{\Psi^+}_\text{T}$ and $\ket{\Psi^+}_\text{C}\ket{11}_\text{T}$, and thus we consider the linear combinations $\ket{E_\pm}$,
\begin{equation}
  \begin{pmatrix}
    \ket{E_+} \\ \ket{E_-}
  \end{pmatrix}
  =
  \begin{pmatrix}
    \cos\vartheta & \sin\vartheta \\
    -\sin\vartheta & \cos\vartheta
  \end{pmatrix}
  \begin{pmatrix}
    \ket{\Psi^+}_\text{C}\ket{11}_\text{T} \\
    \ket{11}_\text{C}\ket{\Psi^+}_\text{T}
  \end{pmatrix}
\end{equation}
which are eigenstates with energies $E_\pm = (J_\text{C} \pm \kappa)/2$, where $\kappa \equiv (1/\Delta) \sqrt{64J^4 + 16J^2 J_\text{C}(J_\text{C} + \Delta) + J_\text{C}^2 \Delta^2 }$, and the mixing angle, $\vartheta$, is defined through
\begin{equation}
    \label{eq:tantheta}
    \tan\vartheta = \frac{2J_\text{C}J}{E_+\Delta + 4J^2} \; .
\end{equation}
Expanding $\ket{11}_\text{C}\ket{01}_\text{T}$ and $\ket{11}_\text{C}\ket{10}_\text{T}$ in eigenstates, the dynamics are
\begin{equation}
  U(t)
  \begin{pmatrix}
    \ket{11}_\text{C} \ket{01}_\text{T} \\ \ket{11}_\text{C} \ket{10}_\text{T}
  \end{pmatrix}
  = \frac{1}{\sqrt 2}
  \begin{pmatrix}
   1 \\ -1
  \end{pmatrix}
  \ket{11}_\text{C} \ket{\Psi^-}_\text{T}
  + \frac{1}{\sqrt 2}
  \begin{pmatrix}
    1 \\ 1
  \end{pmatrix}
  \left( \sin\vartheta e^{-iE_+t} \ket{E_+} + \cos\vartheta e^{-iE_-t} \ket{E_-} \right)
\end{equation}
Since the time-evolution of $\ket{11}_\text{C}\ket{01}_\text{T}$ and $\ket{11}_\text{C}\ket{10}_\text{T}$ only differ by a sign on $\ket{11}_\text{C}\ket{\Psi^-}_\text{T}$, the states have swapped after a time $t$ if the dynamical phases account for this relative sign.
Under the assumption that $\Delta$ is much larger than $J$ and $J_\text{C}$ (all assumed positive for simplicity), we can simplify the expressions for the energies by approximating $\kappa \approx \sqrt{J_\text{C}^2 + 16 J^2 J_\text{C}/\Delta} \approx J_\text{C} + 8 J^2/\Delta$. We see from Eq.~\eqref{eq:tantheta} that $\vartheta \ll 1$. Therefore, the amplitude for the unwanted component $\ket{\Psi^+}_\text{C}\ket{11}_\text{C}$ in the final state scales with
\begin{equation}
    \label{eq:sintheta}
    \sin\vartheta \approx \tan\vartheta \approx \frac{1/4J}{t_g/2\pi + 1/J_\text{C}} \; ,
\end{equation}
which illustrates a trade-off between the gate fidelity and gate time. However, this unwanted state component only leads to small gate infidelites. Ignoring this small effect,
\begin{equation}
  U(t)
  \begin{pmatrix}
    \ket{11}_\text{C} \ket{01}_\text{T} \\ \ket{11}_\text{C} \ket{10}_\text{T}
  \end{pmatrix}
  \approx \frac{1}{\sqrt 2}
  \begin{pmatrix}
   1 \\ -1
  \end{pmatrix}
  \ket{11}_\text{C} \ket{\Psi^-}_\text{T}
  + \frac{1}{\sqrt 2}
  \begin{pmatrix}
    1 \\ 1
  \end{pmatrix}
  e^{i4J^2 t / \Delta} \ket{11}_\text{C}\ket{\Psi^+}_\text{T}
\end{equation}
leading to the desired state swap $\ket{01}_\text{T} \leftrightarrow \ket{10}_\text{T}$ at the gate time $t_g$.

To conclude the discussion of the $U_\text{T}^{11}$ gate of Eq.~\ref{eq:UT11}, we must consider the dynamic evolution of $\ket{11}_\text{C}\ket{11}_\text{T}$ and $\ket{11}_\text{C}\ket{00}_\text{T}$.
The former is a zero-energy eigenstate, cf. Eq.~\eqref{eq:HF_on_1111}, unchanged by time, but the latter is not an eigenstate and mixes with other states. However, comparing Eqs.~\eqref{eq:HF_on_11Psi+} and \eqref{eq:HF_on_1100}, we see that this mixing with unwanted states is essentially the same problem discussed above.
Thus, up to similar small effects, $\ket{11}_\text{C}\ket{00}_\text{T}$ picks up a phase factor of $e^{i4J^2 t_g /\Delta} = -1$ during the gate operation, as desired.
From the numerical simulations of the average fidelity discussed in Section~\ref{sec:simulations}, we find that the average infidelity of the $U_\text{T}^{11}$ gate is well-estimated by $(2J/\Delta)^2 = \pi/(t_g\Delta)$, which is in qualitative agreement with Eq.~\ref{eq:sintheta} for the scaling of the unwanted states' amplitude.

The case where the control qubits are initialized in $\ket{00}_\text{C}$ is analogous. On the other hand, both $\ket{00}_\text{C}$ and $\ket{11}_\text{C}$ couples equivalenly to $\ket{\Psi^+}_\text{C}$, providing two channels for gate infidelities when the control is initialized in $\ket{\Psi^+}_\text{C}$, and hence a larger infidelity.
Numerically, we indeed find that the average infidelity is twice as large, $2\pi/(t_g\Delta)$, for the $U_\text{T}^{\Psi^+}$ compared to $U_\text{T}^{00}$ and $U_\text{T}^{11}$.

\subsection{Equivalent gate diagram for the diamond gate}

\begin{figure}[h]
    \includegraphics[width=.8\columnwidth]{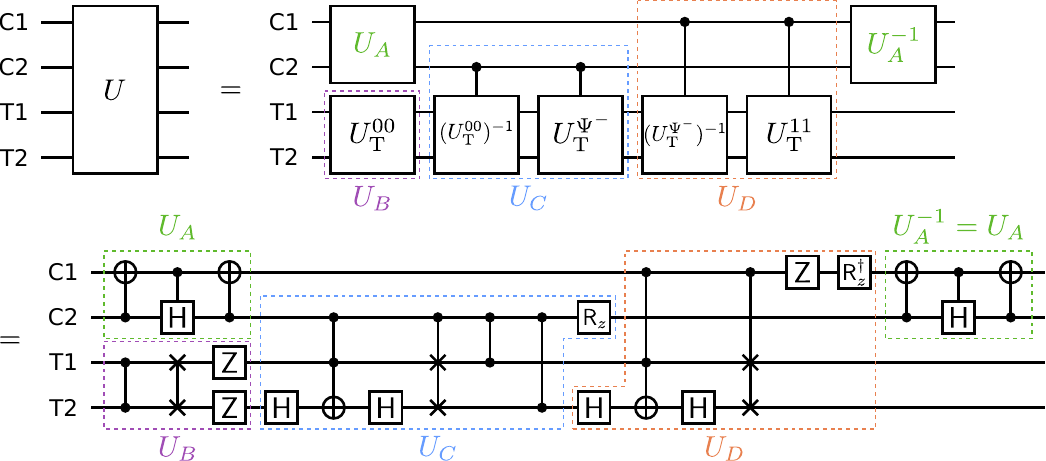}
    \caption{Decomposition of the diamond gate in standard gates from the literature.}
    \label{fig:gate_circuit}
\end{figure}

The four-qubit unitary of the diamond gate of Eq.~\eqref{eq:U}, $U \equiv U(t_g)$, can be expressed in terms of simpler gates in a quantum gate circuit. One way to express the diamond gate is shown in Figure~\ref{fig:gate_circuit}.

This decomposition is found from the following considerations.
We notice that the two-qubit operations performed on the target qubits have simple decompositions in well-known gates, cf. Eqs.~\eqref{eq:UT00}--\eqref{eq:UTmi}, but writing these as conditional operations on the control qubits is not straight-forward due to the Bell states among the control states.
As the first operation in our decomposition, we therefore apply the unitary $U_A$ comprised of two $\textsf{CNOT}$ gates and one controlled-$\textsf{H}$, which maps the control states to the computational basis:
$U_A \cdot \{\ket{00}_\text{C}, \ket{11}_\text{C}, \ket{\Psi^+}_\text{C}, \ket{\Psi^-}_\text{C} \} = \{ \ket{00}_\text{C}, \ket{11}_\text{C}, \ket{10}_\text{C}, -\ket{01}_\text{C} \}$.
The diamond gate leaves the control states unaltered, and we therefore apply the inverse transformation, $U_A^{-1}$, as the last step in the decomposition.
After the application of $U_A$, the two-qubit gates of Eqs.~\eqref{eq:UT00}--\eqref{eq:UTmi} are conditional on the computational control states, making a decomposition much more manageable.

The operations on the target qubits are decomposed in three unitary blocks: $U_B$, $U_C$ and $U_D$.
The first operation, $U_B = U_\text{T}^{00}$, is unconditional on the control qubits, and is the only gate applied when both control qubits are in the $\ket{0}$ state.
Next, if only C2 is in $\ket{1}$ (after $U_A$), the target qubits must be subject to $U_\text{T}^{\Psi^-}$, which we implement in $U_C$.
If both C1 and C2 are in $\ket{1}$, the gate $U_\text{T}^{11}$ must be applied to the target qubits, which is implemented in $U_D$.
Finally, if only C1 is in $\ket{1}$, the gate circuit performs $U_\text{T}^{11} (U_\text{T}^{\Psi^-})^{-1} U_\text{T}^{00} = -\textsf{ZZ} \, e^{-it_g J_\text{C}} = U_\text{T}^{\Psi^+}$, thereby verifying the diamond gate operation.

Expressing these unitary operations in standard gates leads us to the final line in the figure.
Here $\textsf{R}_z = \dyad{0}e^{-i t_g J_\text{C}/2} + \dyad{1}e^{i t_g J_\text{C}/2}$ is $z$-rotation.
This gate diagram can be further decomposed into e.g. $\textsf{CNOT}$ gates and single-qubit rotations.
Using the open-source Python toolbox Qiskit\cite{Qiskit}, we find such a decomposition into 42 $\textsf{CNOT}$s and 49 single-qubit rotations.

\section{Superconducting circuit analysis}
\label{sec:qutrit_analysis}

\begin{figure}[t]
   \includegraphics[width=.5\columnwidth]{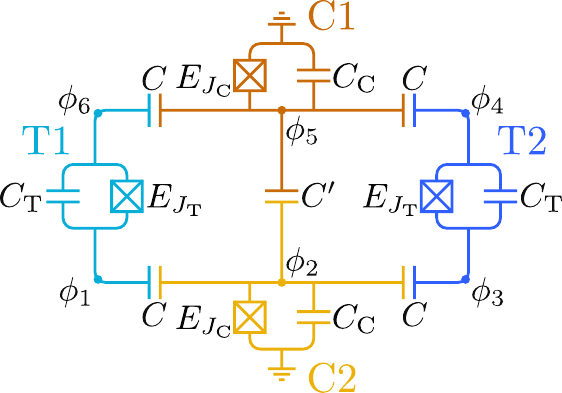}
   \caption{Lumped circuit element diagram of the device from Figure~\ref{fig:superconducting_circuit} with the relevant circuit parameters marked. Node fluxes relative to ground are denoted $\phi_i$ with $i=1,\dots,6$, capacitances are $C$, $C'$, $C_\text{T}$ and $C_\text{C}$, and Josephson energies are $E_{J_\text{T}}$ and $E_{J_\text{C}}$.}
   \label{fig:sc_circuit_analysis}
\end{figure}

In this appendix we analyse the superconducting circuit device of Figure~\ref{fig:superconducting_circuit}, shown as a lumped element diagram in Figure~\ref{fig:sc_circuit_analysis}.
We quantize the circuit using standard techniques~\cite{devoret1995} and truncate each anharmonic oscillator degree of freedom to qutrits (three-level systems), thus arriving at the Hamiltonian~\eqref{eq:qutrit_H0}--\eqref{eq:qutrit_Hint}.
From the qutrit Hamiltonian, the qubit Hamiltonian~\eqref{eq:H0}--\eqref{eq:Hint} follows readily by ignoring all terms involving the second excited transmon states, $\ket{2}$, and ignoring the small crosstalk term, i.e. putting $J_\text{T}=0$.
We study the control qubit subspace in the qutrit model and derive the redefined control state~\ref{eq:tilde11}.
Finally, we derive the optimal crosstalk strength, $J_\text{T}^\text{opt}$ of Eq.~\ref{eq:JTopt}, for countertacting unwanted leakage through the second-excited states.

\subsection{Derivation of the qutrit model Hamiltonian}

The classical Lagrangian for the circuit of Figure~\ref{fig:sc_circuit_analysis} is
\begin{equation}
    \label{eq:Lagrangian}
    \begin{aligned}
        \mathcal{L} ={} &\frac{C_\text{C}}{2} \dot\phi_2^2
            + \frac{C}{2} \left(\dot\phi_3 - \dot\phi_2\right)^2
            + \frac{C_\text{T}}{2} \left(\dot\phi_4 - \dot\phi_3\right)^2
            + \frac{C}{2} \left(\dot\phi_1 - \dot\phi_2\right)^2 \\
            &+ \frac{C_\text{T}}{2} \left(\dot\phi_6 - \dot\phi_1\right)^2
            + \frac{C}{2} \left(\dot\phi_5 - \dot\phi_6\right)^2
            + \frac{C_\text{C}}{2} \dot\phi_5^2
            + \frac{C}{2} \left(\dot\phi_4 - \dot\phi_5\right)^2
            + \frac{C'}{2} \left(\dot\phi_2 - \dot\phi_5\right)^2 \\
            &+ E_{J_\text{C}} \cos\left(\phi_2 \right)
            + E_{J_\text{C}} \cos\left(\phi_5 \right)
            + E_{J_\text{T}} \cos\left(\phi_6 - \phi_1 \right)
            + E_{J_\text{T}} \cos\left(\phi_4 - \phi_3 \right) \; ,
    \end{aligned}
\end{equation}
Here $\phi_i$, for $i=1,\dots 6$, are node fluxes relative to ground at the positions displayed in Figure~\ref{fig:sc_circuit_analysis}. The capacitances and Josephson energies of the model are shown in Figure~\ref{fig:sc_circuit_analysis} at their respective circuit elements.
In the following we used units where the flux quantum is $\Phi_0 = h/(2e) = 2\pi$, in addition to $\hbar = 1$.

We transform the node flux coordinates $\vec{\phi'} = (\phi_1, \phi_2, \phi_3, \phi_4, \phi_5, \phi_6)^T$ using the transformation matrix
\begin{equation}
    T =
    \begin{pmatrix}
        0 & 0 & 0 & 0 & \makebox[1.1em]{$-1$} & 0 \\[.2em]
        0 & \makebox[1.1em]{$1$} & 0 & 0 & 0 & 0 \\[.2em]
        \makebox[1.1em]{$1$} & 0 & 0 & 0 & 0 & \makebox[1.1em]{$-1$} \\[.2em]
        0 & 0 & \makebox[1.1em]{$1$} & \makebox[1.1em]{$-1$} & 0 & 0 \\[.2em]
        \makebox[1.1em]{$1$} & 0 & 0 & 0 & 0 & \makebox[1.1em]{$1$} \\[.2em]
        0 & 0 & \makebox[1.1em]{$1$} & \makebox[1.1em]{$1$} & 0 & 0
    \end{pmatrix} \; ,
\end{equation}
into the new coordinate vector $T \vec{\phi'} \equiv \vec{\phi} = (\phi_\text{C1}, \phi_\text{C2}, \phi_\text{T1}, \phi_\text{T2}, \phi_\text{CM,T1}, \phi_\text{CM,T2})^T$, containing coordinates for each transmon C1, C2, T2, and T2, and two center-of-mass coordinates.

Defining the capacitance matrix
\begin{equation}
    \mathcal{C} =
    \begin{pmatrix}
        C + C_\text{T} & -C & 0 & 0 & 0 & -C_\text{T} \\[.2em]
        -C & C_\text{C} + C' + 2C & -C & 0 & -C' & 0 \\[.2em]
        0 & -C & C + C_\text{T} & -C_\text{T} & 0 & 0 \\[.2em]
        0 & 0 & -C_\text{T} & C + C_\text{T} & -C & 0 \\[.2em]
        0 & -C' & 0 & -C & C_\text{C} + C' + 2C & -C \\[.2em]
        -C_\text{T} & 0 & 0 & 0 & -C & C + C_\text{T}
    \end{pmatrix}
\end{equation}
and the transformed capacitance matrix $K = (T^T)^{-1} \mathcal{C} T^{-1}$, we can express the Lagrangian as
\begin{equation}
    \mathcal{L} = \frac{1}{2} \dot{\vec{\phi}}^T K \dot{\vec{\phi}}
    + E_{J_\text{C}} \cos\left(\phi_\text{C1} \right)
    + E_{J_\text{C}} \cos\left(\phi_\text{C2} \right)
    + E_{J_\text{T}} \cos\left(\phi_\text{T1} \right)
    + E_{J_\text{T}} \cos\left(\phi_\text{T2} \right) \; .
\end{equation}
The generalized momenta (in the new coordinate system) is $\vec{p} = \frac{\partial \mathcal L}{\partial \dot{\vec{\phi}}} = K \dot{\vec{\phi}}$, and hence the classical Hamiltonian is
\begin{equation}
    \label{eq:H_classical_cosines}
    \begin{aligned}
        \mathcal{H} &= \vec{p}^T \dot{\vec{\phi}} - \mathcal L \\
          &= \frac{1}{2} \vec{p}^T K^{-1} \vec{p}
          - E_{J_\text{C}} \cos\left(\phi_\text{C1} \right)
          - E_{J_\text{C}} \cos\left(\phi_\text{C2} \right)
          - E_{J_\text{T}} \cos\left(\phi_\text{T1} \right)
          - E_{J_\text{T}} \cos\left(\phi_\text{T2} \right) \; .
    \end{aligned}
\end{equation}
The capacitance matrix $K$ can be inverted analytically:
\begin{equation}
    K^{-1} =
    \begin{pmatrix}
        8E_{C_\text{C}} & -\mathcal{E}_\text{CC} & \mathcal{E}_\text{CT} & \mathcal{E}_\text{CT} & -\mathcal{E}_\text{C,CM} & -\mathcal{E}_\text{C,CM} \\[.2em]
        -\mathcal{E}_\text{CC} & 8E_{C_\text{C}} & \mathcal{E}_\text{CT} & \mathcal{E}_\text{CT} & \mathcal{E}_\text{C,CM} & \mathcal{E}_\text{C,CM} \\[.2em]
        \mathcal{E}_\text{CT} & \mathcal{E}_\text{CT} & 8E_{C_\text{T}} & \mathcal{E}_\text{TT} & 0 & 0 \\[.2em]
        \mathcal{E}_\text{CT} & \mathcal{E}_\text{CT} & \mathcal{E}_\text{TT} & 8E_{C_\text{T}} & 0 & 0 \\[.2em]
        -\mathcal{E}_\text{C,CM} & \mathcal{E}_\text{C,CM} & 0 & 0 & 8E_{C_\text{CM}} & \mathcal{E}_\text{CM,CM} \\[.2em]
        -\mathcal{E}_\text{C,CM} & \mathcal{E}_\text{C,CM} & 0 & 0 & \mathcal{E}_\text{CM,CM} & 8E_{C_\text{CM}}
    \end{pmatrix} \; ,
\end{equation}
where
\begin{align}
    E_{C_\text{C}} &= \frac{1}{8} \frac{2C_\text{T} (C_\text{C} + C') + C(C_\text{C} + C' + 2C_\text{T})}{C_\text{C}(2C_\text{T}(C_\text{C} + 2C') + C(C_\text{C} + 2C' + 4C_\text{T}))} \; , \\[.2em]
    E_{C_\text{T}} &= \frac{1}{8} \frac{2(C^2 + 2C_\text{T}(C_\text{C} + 2C') + C(C_\text{C} + 2C' + 4C_\text{T}))}{(C+2C_\text{T})(2C_\text{T}(C_\text{C}+2C') + C(C_\text{C} + 2C' + 4C_\text{T}))} \; , \\[.2em]
    E_{C_\text{CM}} &= \frac{1}{8} \frac{2(C+C_\text{C})}{CC_\text{C}} \; , \\[.2em]
    \mathcal{E}_\text{CC} &= \frac{2C'C_\text{T} + C(C' + 2C_\text{T})}{C_\text{C}(2C_\text{T}(C_\text{C} + 2C') + C(C_\text{C} + 2C' + 4C_\text{T}))} \; , \\[.2em]
    \mathcal{E}_\text{CT} &= \frac{C}{2C_\text{T}(C_\text{C} + 2C') + C(C_\text{C} + 2C' + 4C_\text{T})} \; , \\[.2cm]
    \mathcal{E}_\text{TT} &= \frac{2C^2}{(C+2C_\text{T})(2C_\text{T}(C_\text{C} + 2C') + C(C_\text{C} + 2C' + 4C_\text{T}))} \; , \\[.2em]
    \mathcal{E}_\text{C,CM} &= \frac{1}{C_\text{C}} \; , \\[.2em]
    \mathcal{E}_\text{CM,CM} &= \frac{2}{C_\text{C}} \; .
\end{align}

We assume that $C_\text{C}, C_\text{T} \gg C, C'$ which means that each colored circuit area in Figure~\ref{fig:sc_circuit_analysis} may be regarded as a well-defined transmon, and couplings between transmons as perturbations.
In the weak coupling limit, the capacitive energies for the transmons are $E_{C_\text{C}} \approx \frac{1}{8C_\text{C}}$ and $E_{C_\text{T}} \approx \frac{1}{8C_\text{T}}$, while the energy of the center-of-mass degrees of freedom is $E_{C_\text{CM}} \approx \frac{1}{4C} \gg E_{C_\text{C}}, E_{C_\text{T}}$.
In analogy with classical particles, the transmons correspond to pendulums of mass $C_\text{C}$ and $C_\text{T}$, while the center-of-mass degrees of freedom correspond to a very light free particle. We will therefore ignore the center-of-mass degrees of freedom, as their motion will primarily contribute a constant energy shift, which does not affect the dynamics of the transmons.
Notice that the crosstalk coupling between T1 and T2, $\mathcal{E}_\text{TT} \approx \frac{C^2}{2C_\text{T}^2C_\text{C}}$, is suppressed compared to the C1-C2 coupling $\mathcal{E}_\text{CC} \approx \frac{C'+C}{C_\text{C}^2}$ and the control-target coupling $\mathcal{E}_\text{CT} \approx \frac{C}{2C_\text{T} C_\text{C}}$.
This is expected, as the circuit has no direct capacitive coupling between T1 and T2.

Since we intend to operate the transmons near the ground state, we can assume that each transmon is near the vicinity of the potential minimum, thus allowing a fourth order expansion of the cosines in the Hamiltonian~\eqref{eq:H_classical_cosines}. Up to an irrelevant constant energy shift, we arrive at
\begin{equation}
    \label{eq:H_classical_nocosines}
    \mathcal H = \sum_{\nu = \text{C},\text{T}} \sum_{i=1,2} H_{\text{cl}, \nu i}
        + \sum_{i,j = 1,2} \mathcal{E}_\text{CT} \, p_{\text{C} i} p_{\text{T} j}
        - \mathcal{E}_\text{CC} \, p_\text{C1} p_\text{C2}
        + \mathcal{E}_\text{TT} \, p_\text{T1} p_\text{T2} \; ,
\end{equation}
where the Hamiltonian for the non-interacting transmon $\nu i$ is
\begin{equation}
    \mathcal{H}_{\nu i} = 4E_{C_\nu} p_{\nu i}^2 + \frac{1}{2} E_{J_\nu} \phi_{\nu i}^2 - \frac{1}{24} E_{J_\nu} \phi_{\nu i}^4 \; .
\end{equation}
The first two terms in $\mathcal{H}_{\nu i}$ describe a harmonic oscillator, while the last term is a small anharmonic term.

We quantize the system by mapping the classical conjugate coordinates to the quantum operators:
\begin{align}
    \phi_{\nu i} &\mapsto \left( \frac{2E_{C_\nu}}{E_{J_\nu}} \right)^{1/4} (b_{\nu i}^\dagger + b_{\nu i}) \\
    p_{\nu i} &\mapsto i \left( \frac{E_{J_\nu}}{32 E_{C_\nu}} \right)^{1/4} (b_{\nu i}^\dagger - b_{\nu i}) \; ,
\end{align}
for $\nu = \text{C}, \text{T}$ and $i = 1,2$.
Here $b_{\nu i}$ is the usual bosonic annihilation operator, which diagonalizes the harmonic oscillator part of the Hamiltonian, such that the mapping for transmon $\nu i$ to a quantum Hamiltonian is
\begin{equation}
    \mathcal{H}_{\nu i}
    \mapsto
    H_{\nu i} =
    \sqrt{8E_{C_\nu} E_{J_\nu}} \left( b_{\nu i}^\dagger b_{\nu i} + \frac{1}{2} \right)
    - \frac{1}{12} E_{C_\nu} \left( b_{\nu i} + b_{\nu i} \right)^4 \; .
\end{equation}
In the basis of harmonic oscillator states, $\ket{n}_{\nu i}^\text{HO}$ for $n=0,1,2,\dots$, the annihilation operator is $b_{\nu i} = \sum_{n=1}^\infty \sqrt{n} \dyad{n-1}{n}_{\nu i}^\text{HO}$. We assume that the transmons are operated near their ground states, and we therefore truncate each single-transmon Hilbert space to the first three harmonic oscillator states. Up to a constant energy shift, this results in the single-qutrit Hamiltonian,
\begin{equation}
    H_{\nu i}
    \mapsto
    \tilde H_{\nu i} = (\sqrt{8E_{C_\nu} E_{J_\nu}} - E_{C_\nu}) \dyad{1}_{\nu i}^\text{HO}
    + (2\sqrt{8E_{C_\nu} E_{J_\nu}} - 3E_{C_\nu}) \dyad{2}_{\nu i}^\text{HO}
    - \frac{E_{C_\nu}}{\sqrt 2} \dyad{0}{2}_{\nu i}^\text{HO}
    - \frac{E_{C_\nu}}{\sqrt 2} \dyad{2}{0}_{\nu i}^\text{HO} \; .
\end{equation}
We denote qutrit operators with tildes to distinguish them from, e.g., the qubit operators used in Sections~\ref{sec:system}--\ref{sec:simulations}. Notice that inclusion of the third state, $\ket{2}_{\nu, i}^\text{HO}$, introduces mixing terms in the Hamiltonian. Thus, unlike the case of truncation to qubits, where the two lowest harmonic oscillator states become the qubit states, we have to diagonalize the above Hamiltonian. Doing so, we find the qutrit states:
\begin{align}
    \label{eq:qutrit_state0}
    \ket{0}_{\nu i} &= \frac{1}{\sqrt{\frac{1}{2}E_{C_\nu}^2 + \omega_{0,\nu}^2}} \left( \frac{E_{C_\nu}}{\sqrt 2} \ket{0}_{\nu i}^\text{HO} - \omega_{0,\nu} \ket{2}_{\nu i}^\text{HO} \right) \; \\
    \label{eq:qutrit_state1}
    \ket{1}_{\nu i} &= \ket{1}_{\nu i}^\text{HO} \; \\
    \label{eq:qutrit_state2}
    \ket{2}_{\nu i} &= \frac{1}{\sqrt{\frac{1}{2}E_{C_\nu}^2 + \omega_{2,\nu}^2}} \left( -\frac{E_{C_\nu}}{\sqrt 2} \ket{0}_{\nu i}^\text{HO} + \omega_{2,\nu} \ket{2}_{\nu i}^\text{HO} \right) \; ,
\end{align}
where the corresponding energies can be expressed as
\begin{align}
    \omega_{0, \nu} &= \sqrt{\left(\Omega_\nu + \frac{1}{2}\alpha_\nu\right)^2 - \frac{1}{2}\alpha_\nu^2} - \Omega_\nu - \frac{1}{2} \alpha_\nu \; , \\
    \omega_{1, \nu} &= \omega_{0,\nu} + \Omega_\nu \; , \\
    \omega_{2, \nu} &= \omega_{1,\nu} + \Omega_\nu + \alpha_\nu \; .
\end{align}
Here $\Omega_\nu$ is the qubit frequency, i.e. the energy difference between the qubit levels, and $\alpha_\nu$ is the anharmonicity. In terms of circuit parameters, they are given as
\begin{align}
    \Omega_\nu &= \frac{1}{2} E_{C_\nu} + \sqrt{\left(\sqrt{8E_{C_\nu} E_{J_\nu}} - \frac{3}{2} E_{C_\nu}\right)^2 + \frac{1}{2} E_{C_\nu}^2} \; , \\
    \alpha_\nu &= - E_{C_\nu} \; .
\end{align}
In the transmon regime, $E_{C_\nu} \ll E_{J_\nu}$, the anharmonicity is negative and much smaller than the qubit frequency, $-\alpha_\nu \ll \Omega_\nu$.
We can write the qutrit Hamiltonian on the form
\begin{equation}
    \label{eq:single_qutrit_Hamiltonian}
    \tilde H_{\nu i} = -\frac{1}{2} \Omega_\nu \, \tilde \sigma_z^{\nu i} + \frac{1}{2}(\omega_{0,\nu} + \omega_{1,\nu}) \textsf{I}_{\nu i} \; ,
\end{equation}
which is a straightforward generalization of the typical single-qubit Hamiltonian.
Here $\textsf{I}_{\nu i}$ is the identity operator and $\tilde \sigma_z^{\nu i}$ is a generalized Pauli $z$-operator,
\begin{equation}
    \tilde \sigma_z^{\nu i} = \dyad{0}_{\nu i} - \dyad{1}_{\nu i} - \left( 3 + \frac{2\alpha_\nu}{\Omega_\nu} \right) \dyad{2}_{\nu i} \; .
\end{equation}

We now map the interaction terms of the Hamiltonian~\eqref{eq:H_classical_nocosines} to the qutrit model.
Quantization of an interaction term yields
\begin{equation}
    \mathcal{E}_{\nu \mu} \, p_{\nu i} p_{\mu j} \mapsto
    -\mathcal{E}_{\nu \mu} \left( \frac{E_{J_\nu}}{32 E_{C_\nu}} \right)^{1/4} \left( \frac{E_{J_\mu}}{32 E_{C_\mu}} \right)^{1/4}
    (b_{\nu i}^\dagger - b_{\nu i}) (b_{\mu j}^\dagger - b_{\mu j}) \; ,
\end{equation}
where $\nu,\mu \in \{\text{C},\text{T} \}$ and $i,j \in \{1,2 \}$.
Since the operator $b_{\nu i}^\dagger - b_{\nu i} = \sum_{n=1}^\infty \sqrt{n} (\dyad{n}{n-1}_{\nu i}^\text{HO} - \dyad{n-1}{n}_{\nu i}^\text{HO})$,
we expect the coupling between the transmon states $\ket{n-1}_{\nu i}$ and $\ket{n}_{\nu i}$ to be roughly $\sqrt{n}$.
However, if the transmons are initialized in the qubit subspace, spanned by the lowest two states, population of higher order states require higher-order processes and are limited by the number of excitations in the system.
Truncating the Hilbert space to the lowest three harmonic oscillator states maps $b_{\nu i}^\dagger - b_{\nu i} \mapsto - i \tilde \sigma_y^{\nu i}$,
where $\sigma_y^{\nu i}$ is the generalized Pauli $y$-operator on qutrit $\nu i$ defined as
\begin{equation}
    \tilde \sigma_y^{\nu i} = i T_0^\nu \dyad{1}{0}_{\nu i} + i T_2^\nu \dyad{2}{1}_{\nu i} + \text{H.c.} \; ,
\end{equation}
with
\begin{align}
    T_\beta^\nu = \sqrt{2} \, \frac{\omega_{\beta,\nu} - \frac{1}{2} \alpha_\nu }{\sqrt{\omega_{\beta,\nu}^2 + \frac{1}{2} \alpha_\nu^2}} \; , \quad \beta = 0,2 \; .
\end{align}
In the transmon regime $T_0^\nu \approx 1$ and $T_2^\nu \approx \sqrt{2}$, as expected.

Putting these results together and ignoring the constant off-set in Eq.~\eqref{eq:single_qutrit_Hamiltonian}, we find the final qutrit Hamiltonian:
\begin{equation}
    \label{eq:tildeH}
    \tilde H = - \frac{1}{2} \Omega_\text{T} (\tilde \sigma_z^\text{T1} + \tilde \sigma_z^\text{T2})
               - \frac{1}{2} \Omega_\text{C} (\tilde \sigma_z^\text{C1} + \tilde \sigma_z^\text{C2})
               + J_\text{T} \tilde\sigma_y^\text{T1} \tilde\sigma_y^\text{T2}
               + J_\text{C} \tilde\sigma_y^\text{C1} \tilde\sigma_y^\text{C2}
               + J (\tilde\sigma_y^\text{T1} + \tilde\sigma_y^\text{T2}) (\tilde\sigma_y^\text{C1} + \tilde\sigma_y^\text{C2}) \; ,
\end{equation}
which is the sum of $\tilde H_0$ and $\tilde H_\text{int}$ from Eqs.~\eqref{eq:qutrit_H0}--\eqref{eq:qutrit_Hint} in the main text. In terms of circuit parameters, the couplings are
\begin{align}
    J_\text{T} &= \mathcal{E}_\text{TT} \sqrt{\frac{E_{J_\text{T}}}{32 E_{C_\text{T}}}} \; , \\
    J_\text{C} &= - \mathcal{E}_\text{CC} \sqrt{\frac{E_{J_\text{C}}}{32 E_{C_\text{C}}}} \; , \\
    J &= \mathcal{E}_\text{CT} \left( \frac{E_{J_\text{C}}}{32 E_{C_\text{C}}} \right)^{1/4} \left( \frac{E_{J_\text{T}}}{32 E_{C_\text{T}}} \right)^{1/4} \; .
\end{align}

\subsection{Redefinition of the control states}

We wish to understand how the second excited states influence the dynamics of the control states.
In order to make the problem more manageable, we reduce to system size to the control qubits only and consider the effective control Hamiltonian
\begin{equation}
    \tilde H_\text{C} = -\frac{1}{2} \Omega_\text{C} (\tilde \sigma_z^\text{C1} + \tilde \sigma_z^\text{C2})
                        + J_\text{C} \tilde\sigma_y^\text{C1} \tilde\sigma_y^\text{C2} \; ,
\end{equation}
which assumes the target qubits are far detuned from the control qubits, e.g. during state initialization.
We expect this Hamiltonian to dominate the dynamics of the control qubits also when the full four-qubit diamond gate operates.

Let us consider the effect on each control state. First, we see that
\begin{align}
    \tilde H_\text{C} \ket{00}_\text{C}     &= - \Omega_\text{C} \ket{00}_\text{C} - J_\text{C} (T^\text{C}_0)^2 \ket{11}_\text{C} \; , \\
    \label{eq:HC_on_psipm}
    \tilde H_\text{C} \ket{\Psi^\pm}_\text{C} &= \pm J_\text{C} (T_0^\text{C})^2 \ket{\Psi^\pm}_\text{C} - J_\text{C} T_0^\text{C} T_2^\text{C} \frac{1}{\sqrt 2} (\ket{12}_\text{C} \pm \ket{21}_\text{C}) \; .
\end{align}
Since the $\tilde \sigma_y$ operators only couples neighboring energy states, the zero excitation state $\ket{00}_\text{C}$ does not couple to any second excited transmon levels. However, it does couple to $\ket{11}_\text{C}$, but being offset by two excitations, this coupling is energetically suppressed.
The Bell states' couplings to second excited states, i.e. the second term on the right hand side of Eq.~\ref{eq:HC_on_psipm}, can be dismissed with the same argument.
Effectively, the states $\ket{00}_\text{C}$ and $\ket{\Psi^\pm}_\text{C}$ exhibit no dynamics.

This does not hold for the $\ket{11}_\text{C}$ state:
\begin{equation}
    \tilde H_\text{C} \ket{11}_\text{C} = \Omega_\text{C} \ket{11}_\text{C} + J_\text{C} T_0^\text{C} T_2^\text{C} (\ket{02}_\text{C} + \ket{20}_\text{C})
    - J_\text{C} (T_0^\text{C})^2 \ket{00}_\text{C} - J_\text{C} (T_2^\text{C})^2 \ket{22}_\text{C} \; ,
\end{equation}
The last two terms are energetically suppressed, but the coupling to $(\ket{02}_\text{C} \ket{20}_\text{C})/\sqrt 2$ can not be dismissed on this account. Noticing that the Hamiltonian only couples this state back to $\ket{11}_\text{C}$,
\begin{equation}
    \tilde H_\text{C} \frac{1}{\sqrt 2} (\ket{02}_\text{C} + \ket{20}_\text{C}) = (\Omega_\text{C} + \alpha_\text{C}) \frac{1}{\sqrt 2} (\ket{02}_\text{C} + \ket{20}_\text{C})
    + \sqrt{2} J_\text{C} T_0^\text{C} T_2^\text{C} \ket{11}_\text{C} \; ,
\end{equation}
we can diagonalize the Hamiltonian in the subspace spanned by $\ket{11}_\text{C}$ and $(\ket{02}_\text{C} + \ket{20}_\text{C})/\sqrt 2$, yielding two eigenstates.
The redefined control state, $\ket{\tilde{11}}_\text{C}$ of Eq.~\eqref{eq:tilde11}, is the eigenstate that reduces to $\ket{11}_\text{C}$ when $J_\text{C} \rightarrow 0$. When initializing the control in $\ket{\tilde{11}}_\text{C}$ rather that $\ket{11}_\text{C}$, we suppress dynamics in the control state and hence gate infidelity.

\subsection{Engineering crosstalk}

The presence of second excited states in the transmon spectrum allows quantum state transfer between the target qubits, which renders the swap operation unconditional on the control state.
This has essentially the same consequences as a small direct coupling between the target qubits (crosstalk).
Both effects are unavoiable in a superconducting transmon qubit implementation, but we can effectively avoid the state transfer by picking the crosstalk strength such that it cancels the state transfer occuring via the qutrit levels.
The goal of this section is to derive a value of the crosstalk strength which optimally achieves this cancellation.

\begin{figure}
   \includegraphics[width=.5\columnwidth]{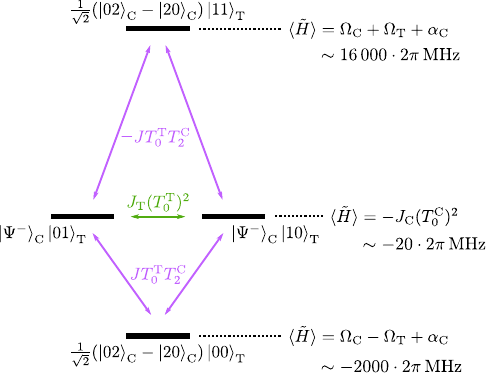}
   \caption{Subspace relevant for the cancellation of unwanted swap between the target qubits. State couplings and their strengths are shown in green and magenta. The energy (expectation value) of the states are indicated, and estimates are given using the same parameters as in the simulations in Section.~\ref{sec:qutrits}.}
   \label{fig:engineering_crosstalk}
\end{figure}

For our analysis here, we assume that the control qubits are initialized in $\ket{\Psi^-}_\text{C}$, which should ideally prevent any dynamics in the system, and that the target qubits are initialized in $\ket{01}_\text{T}$.
The Hamiltonian $\tilde H$ of Eq.~\eqref{eq:tildeH} couples $\ket{\Psi^-}_\text{C} \ket{01}_\text{T}$ to several other states, but we are only interested in processes that contribute significantly to the unwanted swap operation,
\begin{equation}
    \ket{\Psi^-}_\text{C} \ket{01}_\text{T} \rightarrow \ket{\Psi^-}_\text{C} \ket{10}_\text{T} \; .
\end{equation}
Thus we truncate our analysis at second-order contributions, which leaves only the states shown in Figure~\ref{fig:engineering_crosstalk}.
These four states comprise an effective Hilbert space, where the swap can occur as a first-order crosstalk process (green), or a second-order process via the detuned states involving second-excited transmon states (magenta).
The latter processes, being second-order occuring via detuned states, are relatively slow, which means that only a small amount of crosstalk is needed in order to match their transition rates.

In the basis $\{ \ket{\Psi^-}_\text{C}\ket{10}_\text{T}, \frac{1}{\sqrt 2}(\ket{02}_\text{C} - \ket{20}_\text{C}) \ket{00}_\text{T}, \frac{1}{\sqrt 2}(\ket{02}_\text{C} - \ket{20}_\text{C})\ket{11}_\text{T}, \ket{\Psi^-}_\text{C}\ket{01}_\text{T} \}$ the effective Hamiltonian is
\begin{equation}
    \tilde H_\text{eff} =
    \begin{pmatrix}
        0 & \delta & \delta & \kappa \\
        \delta & \Delta_- & 0 & \delta \\
        \delta & 0 & \Delta_+ & \delta \\
        \kappa & \delta & \delta & 0
    \end{pmatrix} \; ,
\end{equation}
with
\begin{align}
    \Delta_\pm &= \Omega_\text{C} \pm \Omega_\text{T} + \alpha_\text{C} + J_\text{C} (T_0^\text{C})^2 \\
    \delta &= J T_0^\text{T} T_2^\text{C} \\
    \kappa &= J_\text{T} (T_0^\text{T})^2 \; .
\end{align}
Thus, the goal is to find the relationship between $\delta$ and $\kappa$ such that the dynamics under the effective Hamiltonain is frozen for the initial state $\ket{\Psi^-}_\text{C}\ket{10}_\text{T}$.
The transition probabilty, which we want to minimize, is
\begin{equation}
    P = \abs{ \bra{10}_\text{T} \bra{\Psi^-}_\text{C} e^{-i \tilde H_\text{eff} t} \ket{\Psi^-}_\text{C} \ket{01}_\text{T} }^2 \; ,
\end{equation}
We consider the problem perturbatively in the effective couplings, writing $\tilde H_\text{eff} = \tilde H_\text{eff,0} + \tilde V_\text{eff}$, with $\tilde H_\text{eff,0}$ being the diagonal and $\tilde V_\text{eff}$ the non-diagonal part of $\tilde H_\text{eff}$.
This enables us to express the time-evolution operator in the interactionpicture, $U_I(t) =  e^{i \tilde H_\text{eff,0} t} \, e^{-i \tilde H_\text{eff} t} \, e^{-i \tilde H_\text{eff,0} t}$, as a Dyson series.
Truncating the perturbative series at second-order contributions,
\begin{equation}
    U_I(t) \approx 1 - i \int_0^t dt' \, e^{i \tilde H_\text{eff,0}t'} \, \tilde V_\text{eff} \, e^{-i \tilde H_\text{eff,0}t'}
    + (-i)^2 \int_0^t dt' \, \int_0^{t'} dt'' \, e^{i \tilde H_\text{eff,0}t'} \, \tilde V_\text{eff} \, e^{-i \tilde H_\text{eff,0}(t'-t'')} \, \tilde V_\text{eff} \, e^{-i \tilde H_\text{eff,0}t''} \; ,
\end{equation}
we find for the transition probability:
\begin{align}
    P &= \abs{ \bra{10}_\text{T} \bra{\Psi^-}_\text{C} U_I(t) \ket{\Psi^-}_\text{C} \ket{01}_\text{T} }^2 \\
      &= \abs{-i \int_0^t dt' \, \kappa - \delta^2 \int_0^t dt' \, \int_0^{t'} dt'' \, \left( e^{i\Delta_-(t''-t')} + e^{i\Delta_+(t''-t')}\right)}^2 \\
      &= \abs{-it\kappa + \delta^2 \left( \frac{i\Delta_- t + e^{-i\Delta_- t} - 1}{\Delta_-^2} + \frac{i\Delta_+ t + e^{-i\Delta_+ t} - 1}{\Delta_+^2}  \right)}^2 \\
      &\approx t^2 \abs{ \kappa - \delta^2 \left( \frac{1}{\Delta_-} + \frac{1}{\Delta_+} \right)}^2 \; ,
\end{align}
where we have ignored the terms of order $\delta^2 / \Delta_\pm^2 \ll 1$ in the last line.
Thus, the condition for a vanishing transition probability is
\begin{equation}
    \kappa = \delta^2 \left( \frac{1}{\Delta_-} + \frac{1}{\Delta_+} \right) \; ,
\end{equation}
which in terms of the crosstalk strength becomes $J_\text{T} = J_\text{T}^\text{opt}$ of Eq.~\eqref{eq:JTopt}.

\twocolumngrid

\bibliography{bibliography}

\end{document}